\documentclass[showpacs, amsmath, amsfonts, amssymb, twocolumn,superscriptaddress,pra]{revtex4}
\usepackage{graphicx}
\usepackage{color}

\usepackage{ulem}  

\begin{document}

\newcommand{\vect}[1]{{\mathrm {\mathbf #1}}} 
\newcommand{\real}[1]{{\mathrm Re}\, #1} 
\newcommand{\imag}[1]{{\mathrm Im}\, #1} 

\title{Analysis of  pattern forming instabilities in an ensemble of two-level atoms optically excited by counter-propagating fields}

\author{ W.~J.~Firth\footnote{To whom correspondence should be addressed.}\email[]{w.j.firth@strath.ac.uk}}
\author{I. Kre\v si\' c}

\affiliation{SUPA and Department of Physics, University of
Strathclyde, 107 Rottenrow East, Glasgow G4 0NG, UK}

\author{G. Labeyrie}

\author{A. Camara}

\affiliation{Institut Non Lin\'{e}aire de Nice, UMR 7335 CNRS, 1361 route des Lucioles, 06560 Valbonne, France}
\author{P. Gomes}

\author{T. Ackemann}
\affiliation{SUPA and Department of Physics, University of
Strathclyde, 107 Rottenrow East, Glasgow G4 0NG, UK}

\date{\today}

\begin{abstract}

We explore various models for the pattern forming instability in a laser-driven cloud of cold two-level atoms with a plane feedback mirror. Focus is on the combined treatment of nonlinear propagation in a diffractively thick medium and the boundary condition given by feedback. The combined presence of purely transverse transmission gratings and reflection gratings on wavelength scale is addressed. Different truncation levels of the Fourier expansion of the dielectric susceptibility in terms of these gratings are discussed and compared to literature. A formalism to calculate the exact solution for the homogenous state in presence of absorption is presented. The relationship between the counterpropagating beam instability and the feedback instability is discussed. Feedback reduces the threshold by a factor of two under optimal conditions. Envelope curves which bound all possible threshold curves for varying mirror distances are calculated. The results are comparing well to experimental results regarding the observed length scales and threshold conditions. It is clarified where the assumption of a diffractively thin medium is justified.

\end{abstract}

\maketitle

\section{Introduction}

\begin{figure}
 \includegraphics[scale=0.9,width=\columnwidth,trim=220 100 190 100,clip]{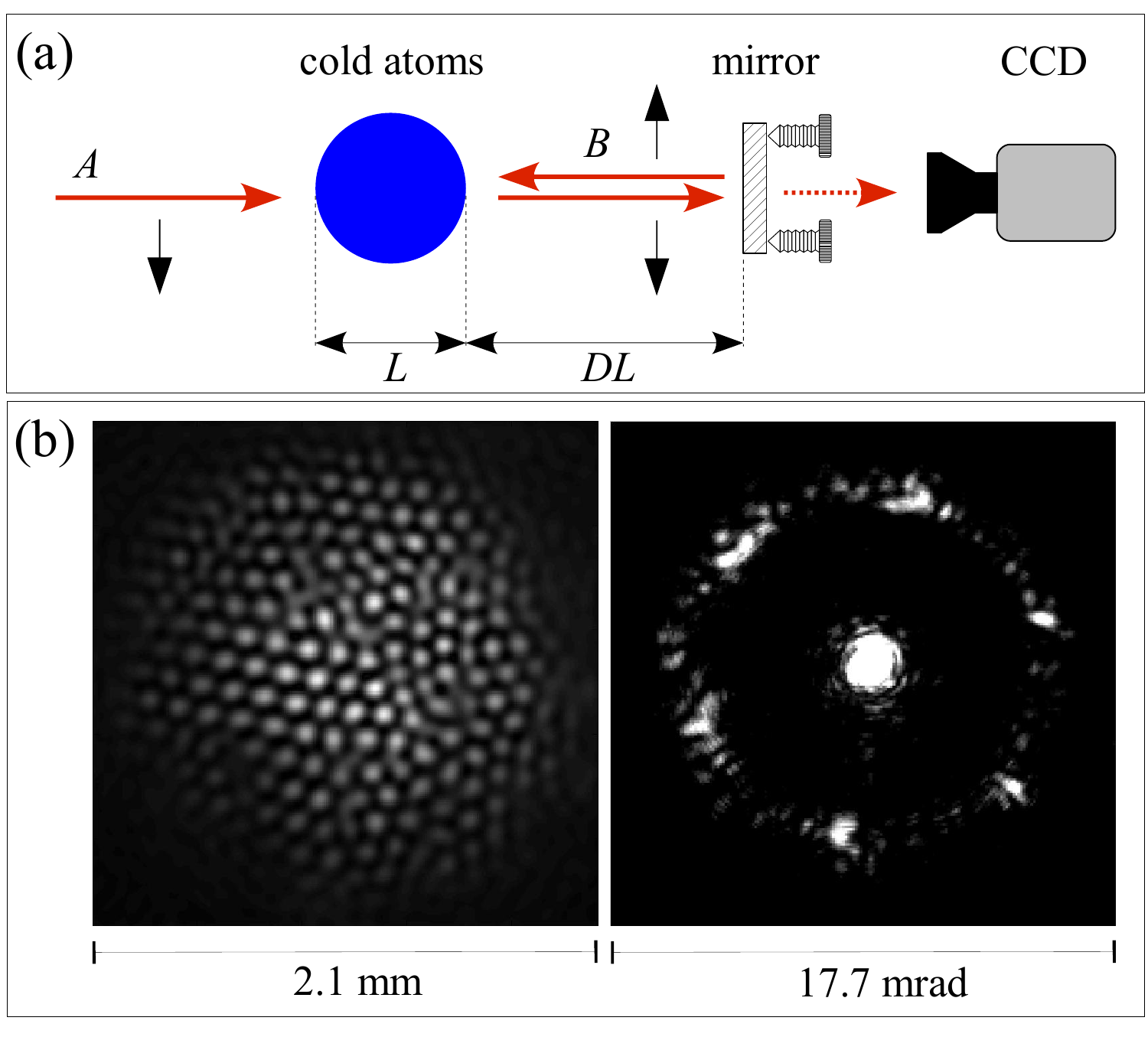}
\caption{ \label{fig:setup}
(Color online)  (a) Experimental SFM
scheme  \cite{Camara2015}: a linearly-polarized light pulse is sent into an atomic cloud; the transmitted beam is retro-reflected by a mirror with an adjustable distance $DL$ beyond the end of the cloud. (b) Typical single-shot light distributions observed in  the transverse instability regime, in the near (left) and far (right) field. Parameters: cloud of $^{87}$Rb atoms at $T=200\,\mu$K driven at a detuning of $\delta =+6.5 \Gamma$ to the $F=2\to F'=3$ transition of the D$_2$-line with an intensity of 0.47~W/cm$^2$, optical density in line center OD$=210$, effective sample size (FWHM of cloud) $L=8.5$~mm.
}
\end{figure}

Optical pattern formation in driven nonlinear media has been studied extensively in late 1980s and 1990s. After observations in four-wave mixing experiments with sodium vapors in counter-propagating (CP) beam configuration reported by Grynberg et. al. \cite{Grynberg1988}, transverse patterns have been observed in liquid crystals \cite{Macdonald1992, Thuring1993,pampaloni94, Vorontsov1995}, thin organic films \cite{Gluckstad1995}, photorefractives \cite{Honda1993,schwab99} and alkali vapors \cite{grynberg94,Ackemann1994,ackemann95b} in the single feedback mirror (SFM) configuration proposed in \cite{firth90a,dalessandro92} (see Fig.~\ref{fig:setup}a for a scheme). Recent years have seen a resurgence of interest in study of transverse self-organization with cold atomic gases in both CP \cite{Greenberg2011, Schmittberger2016} and SFM configurations \cite{labeyrie14, Camara2015} with potential application in condensed matter simulation \cite{Robb2015, CaballeroBenitez2015, Ostermann16}.

Significance of the experiments of Refs.~\cite{labeyrie14,Camara2015} (depicted in Fig. \ref{fig:setup}) is in using optomechanical \cite{Bjorkholm1978,ashkin82} and two-level nonlinearities, respectively. For long pulses ($>10\, \mu$s), with blue detuning, optomechanical density modulation was shown to be dominant in optimum conditions \cite{labeyrie14}. For shorter pulses ($<2\,\mu$s), pattern formation (see Fig.~\ref{fig:setup}b) was found to be consistent with the standard two-level electronic nonlinearity \cite{Camara2015}. Indeed it was already recognized that a two-level instability, though not necessary, was an efficient seed for the optomechanical patterns \cite{labeyrie14}. Here we consider the theory of this two-level instability.


A particularly simple model of the SFM configuration, for a diffractively-thin slice of Kerr medum, was analyzed by Firth \cite{firth90a}. However, as was highlighted in Ref. \cite{Honda1996} and later in \cite{labeyrie14,Camara2015}, the full analysis of pattern properties for small mirror distances demands a "thick-medium" approach, i.e.\ inclusion of diffraction within the non-linear medium. The requisite theory is closely related to that used to analyze pattern formation in a mirrorless thick-medium (slab) with two counterpropagating input fields. Such CP systems have been analyzed for Kerr media by by Firth et al \cite{firth90b} and Geddes et al \cite{Geddes1994}. For a two-level CP system, Muradyan et al  \cite{Muradyan2005}, in an extended abstract for NLGW 2005, describe pattern formation in cold atoms with counter-propagating fields, including both electronic and optomechanical mechanisms, the latter in a molasses model. These works \cite{firth90b,Geddes1994,Muradyan2005} provide the main theoretical background to the present paper, though mention should be made of early analysis \cite{LeBerre1991} aimed at modeling oscillatory spatial instabilities observed in a (hot) sodium vapor SFM experiment \cite{Giusfredi1988}. The present paper concentrates at the modelling of the simultaneous presence of transmission (purely transverse gratings resulting from the interference of the pump with copropagating sidebands) and reflection gratings (wavelength scaled gratings which result from the interference of counterpropagating beams) in presence of the feedback mirror, whereas earlier treatments only utilized pure transmission gratings \cite{firth90a,dalessandro92,ackemann95b}. For the analysis of photorefractive experiments two-beam coupling via pure reflection gratings were considered  \cite{Honda1996}.


As in the Muradyan model (MM) \cite{Muradyan2005}, we use a time-independent susceptibility approach to the two-level nonlinearity. This precludes consideration of growth rates or oscillatory instabilities \cite{LeBerre1991}, but leads to reasonably tractable and transparent models which allow the parameter dependences of pattern thresholds to be investigated. We include absorption, so as to allow for arbitrary atom-field detunings. We also consider the inclusion of reflection-grating effects at different orders (MM include such effects, but only at lowest order). This analysis is then applied to the calculation of thresholds for transverse instability in various thick-medium models. These include the Kerr limit, used for the thick-medium calculations presented in Fig.\ 3B of  \cite{labeyrie14}. In \cite{Camara2015} preliminary two-level results were presented for two cases: quasi-Kerr (i.e. large detuning, neglecting absorption, but not saturation of the refractive nonlinearity) for the pattern size vs mirror distance; and absorptive thin-slice for the threshold vs atomic detuning. The theory behind all these models, as well as additional results for these and related models, will be presented.

As well as elaborating previous preliminary results and presenting generalizations of previous Kerr threshold formulae, we mention two useful and general results which emerge. First, the SFM threshold curve of intensity vs diffraction parameter, for a Kerr-like medium with its feedback mirror directly at its output, coincides with the threshold for the CP instability in a medium of twice the length. This might seem obvious from a 'mirror-image' picture, but there's a twist. The CP thresholds are actually described by two separate but intertwined curves, e.g.\ \cite{Geddes1994}, but only one of these corresponds  to a SFM configuration, for symmetry reasons. Secondly, each member of the family of such threshold curves generated by varying the mirror distance is tangent to an envelope curve, which can be analytically calculated in many cases. This gives useful insight into the mirror-distance dependence of pattern scales, but also enables a quantitative examination of the thin-slice limit, in which diffraction within the medium is neglected. It turns out that  the thin-medium approximation works only at zero order, i.e.\ the threshold at large mirror distance is linear, not quadratic or higher, in $1/D$,
where $L$ is the medium thickness and $d=DL$ the mirror distance.
%

%
%

\section{System and Model}
\label{sec_model}

As in  \cite{Muradyan2005}, we consider the counter-propagating fields $A$ (forward field) and $B$ (backward field, see Fig.~\ref{fig:setup}a) to be coupled by a nonlinear susceptibility

\begin{equation}
\chi_{NL}= - \frac{6\pi}{k_0^3} n_a \frac{2\delta/\Gamma -i}{1+4\delta^2/\Gamma^2}\frac{1}{1+I/I_{s\delta}}
\label{chiNL}
\end{equation}
Here $n_a$ is the atomic density (considered constant here). $I$ is the intensity, which will be a
standing wave: $I/I_{s\delta} = |A e^{ikz} + B e^{-ikz}|^2$. We can conveniently rewrite (\ref{chiNL}) as
\begin{equation}
\chi_{NL}= \chi_l \frac{1}{1+I/I_{s\delta}}
\label{chisat}
\end{equation}
where $\chi_l$ is the linear susceptibility (and is complex, though absorption is neglected in the MM model, making the system Kerr-like).

The next step is to expand the nonlinear factor in a  Fourier series:

\begin{equation}
\frac{1}{1+I/I_{s\delta}}= \sigma_0 +\sigma_+ e^{2ikz} +\sigma_-e^{-2ikz} + h.o.t.
\label{Igrat}
\end{equation}

The higher-order  terms  do not lead to any phase matched couplings, and so can reasonably be neglected whatever the intensity. The coefficients $\sigma_{\pm}$ evidently describe a $2k$ longitudinal modulation of the susceptibility, i.e.\ a reflection grating, which will  scatter the forward field into the backward one and vice versa.

The field equations (M3) of \cite{Muradyan2005} can then be written as

\begin{equation}
\label{modeleqs}   \left \{
\begin{array}{l}
\frac{\partial A}{\partial z} - \frac{ i}{2k}\nabla^2_\perp A =i \frac{k}{2}\chi_l(\sigma_0 A +\sigma_+ B) ,\\
\\ \frac{\partial B}{\partial z} + \frac{ i}{2k}\nabla^2_\perp B = -i \frac{k}{2}\chi_l(\sigma_- A +\sigma_0 B) \\
\end{array} \right.
\end{equation}

To calculate $\sigma_{0,\pm}$, we write the exact expansion of the saturation term (\ref{Igrat}) as
\begin{equation}
\frac{1}{1+I/I_{s\delta}}= \frac{1}{1+a^2+b^2} (1+r(e_{+} + e_{+}^{*}))^{-1}
\label{gratexp}
\end{equation}

where $a=|A|$, $b=|B|$,  $e_{+} = e^{2ikz} e^{i(\theta_A-\theta_B)}$, with
$\theta_{A,B} = arg(A,B)$.  We have introduced a coupling parameter $r=hab/(1+a^2+b^2)$, where the "grating parameter" $h$ \cite{firth90b} has been introduced to allow consistent consideration of the cases of no reflection grating ($h=0$), and of a full grating ($h=1$). In the  former case $\sigma_{\pm}=0$, which would correspond to the standing-wave modulation of the susceptibility being washed out by drift or diffusion. Partial wash-out could be accommodated by intermediate values of $h$, but would need some associated physical justification. The MM model includes the full grating, so corresponds to $h=1$.

The series expansion of $(1+r(e_{+} + e_{+}^{*}))^{-1}$ is always convergent, because $r< 1/2$. Even terms contribute to $\sigma_0$, odd terms to $\sigma_{\pm}$. Using the binomial theorem, we find
\begin{equation}
\label{sigmas}   \left \{
\begin{array}{l}
(1+a^2+b^2)\sigma_0 = 1 +2r^2 +6r^4 + 20r^6 + ...\\
\\ (1+a^2+b^2)\sigma_+ = - e^{i(\theta_A-\theta_B)}(r + 3r^3 + 10r^5 + ...) \\
\end{array} \right.
\end{equation}
with $\sigma_- =  \sigma_+^{*}$.

Inserting these expressions into (\ref{modeleqs}) gives
\begin{equation}
\label{propeqs}   \left \{
\begin{array}{l}
\frac{\partial A}{\partial z} - \frac{ i}{2k}\nabla^2_\perp A =i \frac{k}{2}\chi_lA(\frac{(1+2r^2 + ...) -(br/a)(1+3r^2 + ...)}{1+a^2+b^2}) ,\\
\\ \frac{\partial B}{\partial z} + \frac{ i}{2k}\nabla^2_\perp B = -i \frac{k}{2}\chi_l B(\frac{(1+2r^2 + ...) -(ar/b)(1+3r^2 + ...)}{1+a^2+b^2}) \\
\end{array} \right.
\end{equation}
Note that both sums are positive definite, so higher-order terms reduce the saturation (first term), but increase the strength of the cross-coupling (second term). To lowest order (i.e. cubic nonlinearity), the brackets become $(1-a^2-(1+h)b^2)$ and $(1-(1+h)a^2-b^2)$ for the $A$ and $B$ equations respectively, showing the expected factor of two enhancement of the cross-coupling due to the grating when $h=1$. In the absence of the grating $r=0$, and the bracketed expressions reduce to $(1+s)^{-1}$ in both cases, where $s=a^2+b^2$ is the usual saturation parameter.

The MM model effectively truncates the series expansions in (\ref{propeqs}) at the first term, eventually leading to their equation (M8) for the "transverse eigenvalues"  \cite{Muradyan2005}, which include saturation denominators $\sim(1+s)^{-1}$. However, because $r^2 \sim s^2$, the terms neglected in the MM model are of the same order as the terms $\sim s^5$ which saturate the cubic nonlinearity. For $s= 0.4$ (the value in Fig. 1 of \cite{Muradyan2005}), $r^2$ is only about 0.02, so its neglect is not especially serious in that case.

The series in  (\ref{propeqs}) can be summed. In fact several papers, going back to the 1970s, have obtained analytic solutions to the system (\ref{modeleqs}), or closely equivalent systems (in the plane-wave limit). For our purposes, the papers of van Wonderen et al  \cite{vanW1989,vanWonderen91}, who were addressing optical bistability in a Fabry-Perot cavity, are most directly relevant, and underpin the analytic zero-order (no diffraction) solution obtained in the next section.

Summing the series and  combining both terms leads to a set of field evolution equations:

\begin{equation}
\label{exacteqs}   \left \{
\begin{array}{l}
\frac{\partial A}{\partial z} - \frac{ i}{2k}\nabla^2_\perp A =i \frac{k}{2}\chi_lA(1-\frac{1-2a^2h/(1+a^2+b^2)}{(1-4r^2)^{\frac{1}{2}}})/2a^2h ,\\
\\ \frac{\partial B}{\partial z} + \frac{ i}{2k}\nabla^2_\perp B = -i \frac{k}{2}\chi_l B(1-\frac{1-2b^2h/(1+a^2+b^2)}{(1-4r^2)^{\frac{1}{2}}})/2b^2h \\
\end{array} \right.
\end{equation}
In the limit of no grating, $h,r \to 0$, both brackets reduce to the expected saturation denominator.

For finite $h$, there is explicit nonreciprocity, since the susceptibilities for A and B are different, because of the susceptibility grating. However, the amplitudes $A$ and $B$ are slowly varying in $z$, allowing the propagation in the medium to be approximated by comparitively few longitudinal spatial steps.

In all the cases discussed above, we can write the propagation equations in the form

\begin{equation}
\label{transeqs}   \left \{
\begin{array}{l}
\frac{\partial A}{\partial z} - \frac{ i}{2k}\nabla^2_\perp A =-\frac{\alpha_l}{2}(1+i\Delta) F(a^2,b^2)A ,\\
\\ \frac{\partial B}{\partial z} + \frac{ i}{2k}\nabla^2_\perp B = \frac{\alpha_l}{2}(1+i\Delta) F(b^2,a^2)B \\
\end{array} \right.
\end{equation}
where $\alpha_l$ is the linear absorption coefficient, $\Delta (= 2\delta/\Gamma)$ is the scaled detuning, and the function $F$ describes the nonlinearity of the atomic susceptibility, as modelled by
e.g. (\ref{propeqs}) or (\ref{exacteqs}), by the cubic  ($\chi^{(3)}$) approximation, or some other model.
By definition, $F(0,0) =1$, but $F(a^2,b^2) \ne F(b^2,a^2) $ in general, because of non-reciprocity due to standing-wave effects. The cubic model ($F(a^2,b^2) = 1-a^2 -(1+h)b^2$) is the simplest example, explicitly non-reciprocal if $h \ne 0$.

\begin{figure}
\centering%
\includegraphics[width=\columnwidth]{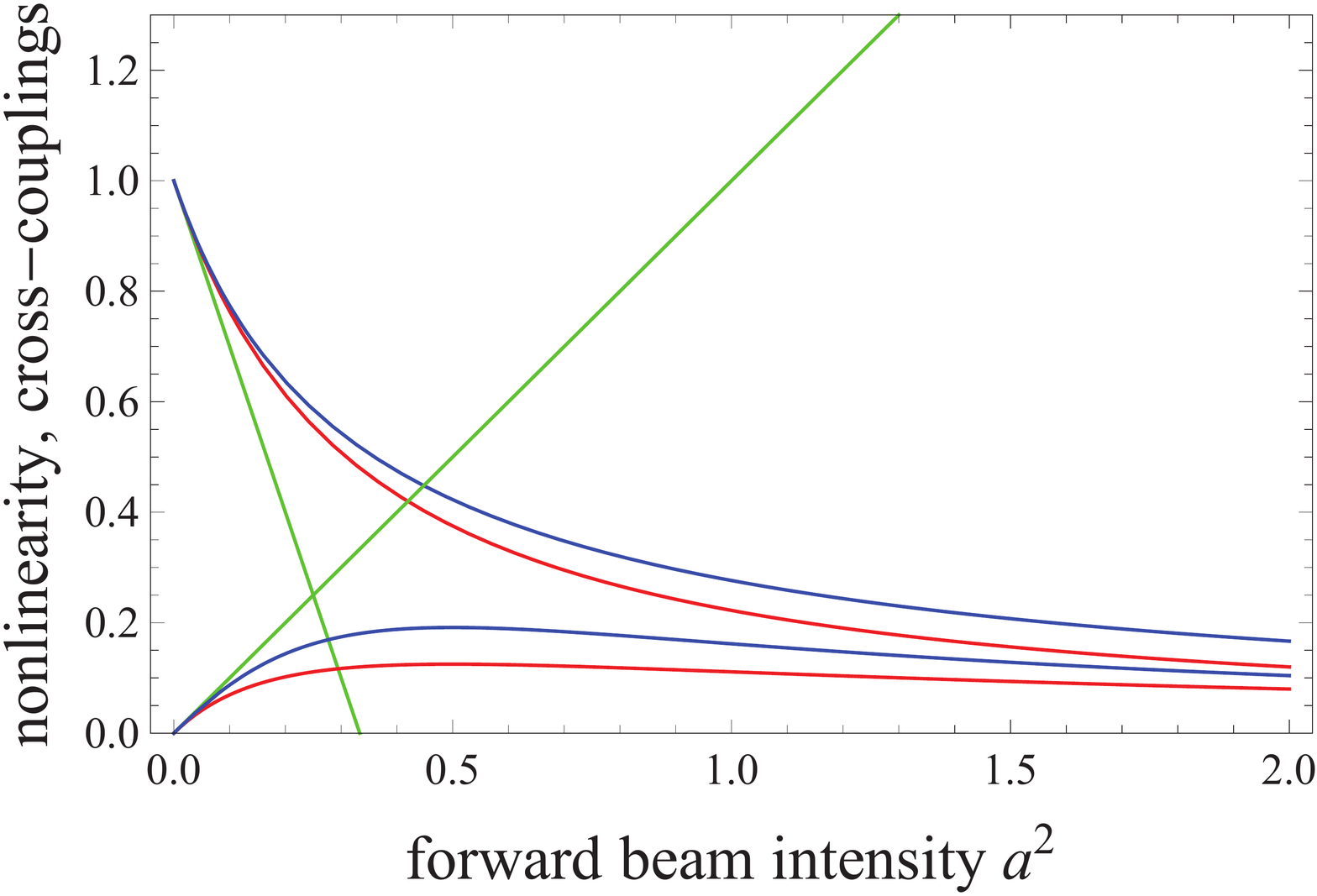}
 \caption{ (color online) Susceptibility and cross-coupling functions against $a^2$ for the equal-intensity case $a=b$, with $h=1$, i.e. full grating. Green: cubic approximation, $F =1-3a^2$. Red: MM model: (\ref{propeqs}) with series truncated at $1$. Blue: full model, no truncation. Susceptibility curves start from $1.0$ at $a=0$, cross-coupling from zero.}

 \label{fig:suscep}
 \end{figure}

Fig. \ref{fig:suscep} illustrates the intensity dependence of the susceptibility and cross-coupling for $h=1$, for the cubic, MM and full models. The cubic (i.e. $\chi^{(3)}$) model evidently has a very limited range of validity, whereas the MM and full models are broadly similar over a broad range, though quantitatively distinct.

\section{Zero-order equations and solutions}
To find the pattern-formation thresholds, we first drop diffraction, and solve the plane-wave, zero-order problem in which $A,B$ depend on $z$ alone. For convenience we set $|A(z)|^2 = p(z)$ and $|B(z)|^2 = q(z)$, and scale $z$ to the medium length $L$. From (\ref{transeqs}) it follows that the plane-wave intensities $p(z), q(z)$ obey the real equations:
\begin{equation}
\label{abzero}   \left \{
\begin{array}{l}
\frac{dp}{d z}  =-\alpha_l L F(p,q)p ,\\
\\ \frac{dq}{d z}  =\alpha_l L F(q,p)q \\
\\
\end{array} \right.
\end{equation}
leading to the expected exponential absorption of the intensities in the linear limit.

We define the input intensity $p(0)=p_0$ and transmitted intensity $p(1)=p_1$, and  similarly $q(0)=q_0$, $q(1)=q_1$. In the SFM  configuration $q_1 = R p_1$, where $R$ is the mirror reflection coefficient, whereas in the CP problem we usually have $q_1 = p_0$.

We now solve (\ref{abzero}) for three different models: no grating $(h=0)$; the MM model ($h=1$, but truncated summations); and the full-grating $h=1$ model based on (\ref{exacteqs}).

For $h=0$, $F=1/(1+s)=1/(1+p+q)$ is symmetric in its arguments, and it follows that the product of the counter-propagating intensities (and indeed of the fields, $AB$) is independent of $z$, simplifying the analysis.  We set $p(z)q(z) = K$, where $K$ is constant, and thus  $K = p_1q_1 = Rp_1^2 $ for a feedback mirror of reflectivity $R$. It follows that the backward intensity $q(z)$ is given by $K/p(z)$, enabling the first member of (\ref{abzero}) to be written in terms of $p(z)$ alone. It can then be integrated analytically, giving
\begin{equation}
\label{p_nograt}
ln(p/p_0)+p-K/p - p_0 + K/p_0 +\alpha_lLz = 0 ,
\end{equation}
and hence, for the transmitted power $p_1$ (using the explicit value of $K$):
\begin{equation}
\label{p_1nograt}   \\
ln(p_1/p_0)+(1-R)p_1=p_0-Rp_1^2/p_0 -\alpha_lL .\\
\\
\end{equation}

For $h=1$, $F(p,q)=(1+p)/(1+s)^2$ in the MM model. We can again find a propagation constant, in this case given by $K=pq/(1+s)$, again leading to a an integrable first-order equation in $p(z)$ alone:
\begin{equation}
\label{dpdz_grat}
\frac{p(1+p)}{(p-K)^2}  \frac{dp}{d z}  =-\alpha_l L.
\end{equation}
This leads, for the transmitted power $p_1$, to
\begin{equation}
\label{p_1grat}   \\
H(p_1,K)-H(p_0,K) +\alpha_lL =0.\\
\\
\end{equation}
where $H(p,K)= p +(2K+1)ln(p-K) - \frac{K(K+1)}{p-K}$.

Finally, it turns out that the all-grating system given by (\ref{exacteqs}) also possesses a propagation constant, given by $K = W(z) - s(z)$, where $ W(z) = (1+2s+\xi^2)^{\frac{1}{2}}$, and $\xi(z) = p(z) - q(z)$. Essentially the same conservation law was noted by Van Wonderen et al in the context of optical bistability in a Fabry-Perot  resonator \cite{vanW1989}, for which the propagation equations are identical to the present case, though the boundary conditions are different.

In terms of $W, s,\xi$ the all-grating function $F_{all} (p,q)$ becomes
$F_{all} = (1+(\xi-1)/W)/(s+\xi)$, with its transpose $F_{all} (q,p)$ obtained by $\xi  \rightarrow -\xi$. Recasting equations (\ref{abzero}), it turns out that the propagation equations for $s$ and $\xi$ take a fairly simple form:
\begin{equation}
\label{sandxi}   \left \{
\begin{array}{l}
\frac{ds}{d z}  = - \alpha_l \xi/W ,\\
\\ \frac{d\xi}{d z}  = - \alpha_l (1-1/W) \\
\\
\end{array} \right.
\end{equation}
from which one easily deduces $dW/dz =ds/dz$, and thus the constancy of $K = W(z) -s(z)$. One can then obtain an integrable differential equation in just one variable. For example, by using the definition of $W$ and of $K$ to express $W$ in terms of $K$ and $\xi$, the second of equations (\ref{sandxi}) is easily integrated to yield:
\begin{equation}
\label{xi_all}   \\
\xi + ln(\xi + (\xi^2+2-2K)^{\frac{1}{2}}) +\alpha_l L z = const.\\
\\
\end{equation}
For the important case $R=1$, we have $s_1 = 2p_1$, $\xi_1 = 0$, hence $W_1 = (1+4p_1)^\frac{1}{2}$ and thus $K = (1+4p_1)^\frac{1}{2} -4p_1$. Using this data in (\ref{xi_all}) yields an implicit expression for $\xi_0$ in terms of $K$ (and thus $p_1$):
\begin{equation}
\label{xi_0}   \\
\xi_0 + ln(\xi_0 + (\xi_0^2+2-2K)^{\frac{1}{2}})  -{\frac{1}{2}} ln(2-2K) = \alpha_l L .\\
\\
\end{equation}
Given $\xi_0$, it is straightforward to calculate $W_0$ and $s_0$, and thus the input intensity $p_0$ and
the backward output intensity $q_0$, all in terms of the given transmitted intensity $p_1$, thus completing the solution of the plane-wave problem for the all-gratings model.

A problem with the MM model arises as the tuning $\Delta$ approaches resonance. It turns out that the transmission determined from (\ref{p_1grat}) shows "bistability", i.e. the output $p_1$ is not a single-valued function of the input $p_0$, if the optical density is high enough. This is surprising and counterintuitive, and turns out to be a flaw in the model: including more terms in the series expansion (\ref{sigmas}) eventually makes $p_1$ single-valued. In particular the all-gratings formula (\ref{xi_all}) and its $R=1$ sub-case (\ref{xi_0}) give single-valued transmission characteristics.


 \section{Transverse perturbations}

 We now assume that a  solution has been found for the plane wave case $A=A_0(z)$, $B=B_0(z)$, obeying appropriate longitudinal boundary conditions. This solution may be numerical, or a solution to some special-case or approximate version of  (\ref{propeqs}). We now turn our attention to the stability of such a plane wave solution against transverse perturbations.
We suppose that the solution of (\ref{abzero}), subject to the appropriate boundary conditions, is known, and consider transverse perturbations of the form $A=A_0(1+f)$, $B=B_0(1+g)$,
where $\nabla^2_\perp(f,g)=-Q^2(f,g)$, i.e. the transverse perturbation has wave vector $Q$, corresponding to a diffraction angle $Q/k$ in the far field. Assuming $|f|,|g| << 1$, we obtain the linearised propagation equations:

\begin{equation}
\label{transpert}   \left \{
\begin{array}{l}
\frac{df}{dz} + \frac{ iQ^2}{2k}f = -\alpha_l(1+i\Delta)
(F_{11}f'+F_{12}g') ,\\
\frac{dg}{dz} - \frac{ iQ^2}{2k}g =  \alpha_l (1+i\Delta)(F_{21}f'+F_{22}g') \\
\end{array} \right.
\end{equation}
Here $f = f' +i f''$, $g=g' + ig''$, and , and the real quantities  $F_{ij}$ are  defined as $F_{11}=p\frac{\partial F(p,q)}{\partial p}$,
$F_{12}=q\frac{\partial F(p,q)}{\partial q}$,
$F_{21}=p\frac{\partial F(q,p)}{\partial p}$,
$F_{22}=q\frac{\partial F(q,p)}{\partial q}$.

We assume that the fields are time-independent, adequate to calculate the threshold of a zero-frequency pattern-forming (Turing) instability at wavevector $Q$. To find Hopf instabilities, or to properly account for dynamical behavior of the field-atom system, we would have to start from the Maxwell-Bloch equations, rather than our susceptibility model. It is worth mentioning that van Wonderen and Suttorp, in a later paper on dispersive optical bistability \cite{vanWonderen91}, perform a perturbation analysis of the full Maxwell-Bloch equations with all grating orders included. The resulting model is very involved, and beyond our present scope. Meantime, we are content to address the Turing pattern formation problem.

\section{Quasi-Kerr case}

Let's begin with the case of large detuning, where the absorption is negligible. The linear absorption coefficient can be written as $\alpha_l = \alpha_0/(1+\Delta^2)$, where $\alpha_0$ is the
on-resonance absorption. Formally, as a quasi-Kerr model, we suppose that $|\Delta|$ is large enough that $\alpha_l L$ can be neglected, but with
$\alpha_l \Delta L$ finite, so that the nonlinearity is purely refractive. For example, recent experiments  \cite{labeyrie14,Camara2015}  employed optical densities $\alpha_0 L$ of order 100, so neglect of absorption is reasonable  for $|\Delta| \sim 20$, which is at the high end of the experimental range. With this assumption,  the forward and backward intensities $p,q$ can be considered constant, and so are the  $F_{ij}$. For feedback mirror boundary conditions, we have $q = R p$, where $R$ is the mirror reflectivity, while for the CP problem $q=p$ if the system is symmetrically pumped.

 Following \cite{firth90b}, we set $\theta=Q^2L/2k$, and recast equations (\ref{transpert}) in this quasi-Kerr limit as

\begin{equation}
\label{transpertK}   \left \{
\begin{array}{l}
\frac{df}{dz} = -i\theta f -i\alpha_l L\Delta(F_{11} f'+ F_{12}g') ,\\
\frac{dg}{dz} = i\theta g + i\alpha_l L\Delta(F_{21}f'+F_{22}g') \\
\end{array} \right.
\end{equation}

It is convenient to define a $2 \times 2$ matrix $\hat{F}$ formed from the $F_{ij}$. The nonlinearly-driven terms in equations (\ref{transpertK}) are pure imaginary, as for Kerr media, because of our assumption on $\Delta$. Importantly, we have not imposed any restrictions on the magnitude of the intensities. If the linear absorption is small, the saturated absorption is even smaller, so our approximation becomes better, not worse, for high intensity. Nor is there any restriction on the form of $F$, so that we can examine and compare different models of nonlinearity and of standing-wave response.

 In order to align with previous work on the CP Kerr case, we first consider the symmetric equal intensity case ($p=q$), for which
$F_{11}=F_{22}=F_{sym}$ and $F_{12}=F_{21}= GF_{sym}$. Both $F_{sym}$ and  $G$ are in general functions of $s=2p$, but are independent of $z$. Thus for any given input(s) (\ref{transpertK}) are formally equivalent to a corresponding Kerr problem, with renormalized intensity and grating factor, and can be solved by the same methods.

To develop the Kerr analogy further, we can write

\begin{eqnarray}
\label{Fsym}
& & \hat{F}_{sym}= F_{sym}  \left(
\begin{array}{cccc}
 1 & G \\
G & 1
\end{array}
\right). \nonumber
\end{eqnarray}
The eigenvalues of $ \hat{F}_{sym}$ are simply given by $F_{sym}(1 \pm G)$, with corresponding eigenvectors proportional to [$1, \pm 1$].\\

We now define $\psi_{1,2}^2 = \theta (\theta +\kappa \phi_{1,2})$, where the effective Kerr coefficient 
$\kappa = \alpha_l L \Delta$. 
($\phi_1, \phi_2$) are the eigenvalues of $\hat{F}$,
chosen such that
($\phi_1, \phi_2) \to F_{sym}(1-G,1+G)$
as $q \to p$. This ensures that  $\psi_{1,2}$  coincide exactly with the quantities $\psi_{1,2}$ used in \cite{firth90b,Geddes1994} in analyzing  the Kerr CP case. It follows that the analysis and results established in these papers for the symmetrically-pumped CP Kerr problem extend to the present quasi-Kerr case, in which both the strength of the nonlinearity and of the grating-coupling $G$ can be intensity dependent (see Appendix for details).
Hence the quasi-Kerr CP  threshold condition is given by the expression familiar from, e.g., \cite{Geddes1994}: \\
\begin{equation}
\label{2LSslab}
2+2cos\psi_1 cos\psi_2 +\left(\frac{\psi_1}{\psi_2} +\frac{\psi_2}{\psi_1} \right)sin\psi_1 sin\psi_2 = 0.
\end{equation}
\\

While this expression is indeed familiar for a Kerr medium, our discussion shows that it applies much more generally, i.e. to any medium (including saturating media) which can be described by a nonlinearity function of the form $F(p,q)$, subject to absorption being negligible. Muradyan et al \cite{Muradyan2005} implied this result in the context of the MM model extended to include some optomechancal effects, but did not explicitly demonstrate it. \\

We now present the explicit forms of the matrix $\hat{F}$ for various models of interest here. For the Kerr case, we have
\begin{equation}
\label{FKerr}
\hat{F}_{kerr}= -
\left(
\begin{array}{cccc}
 p & (1+h)q \\
(1+h)p & q
\end{array}
\right).
\end{equation}
For $p=q$ this leads to $F_{sym} = -p$ and $G=1+h$ as expected.

For the MM model, we obtain
\\
\begin{eqnarray}
\label{Fmur}
& & \hat{F}_{MM}= - \frac{1}{(1+s)^3}\\
& & \left(
\begin{array}{cccc}
 p(1+s)-2hpq & (1+h)q(1+s) -2hq^2 \\
(1+h)p(1+s)-2hp^2 & q(1+s) -2hpq
\end{array}
\right). \nonumber
\end{eqnarray}
For $p=q=s/2$ and $h=1$ the above expression for $\hat{F}_{MM}$ leads to $F_{sym} = - \frac{p}{(1+s)^3}$, while we find an intensity-dependent grating factor  $G=2+s$. This differs from the results of \cite{Muradyan2005}, wherein the given formulae imply  $G=2$. 

The general (all grating terms) function $F$ given in (\ref{exacteqs}) also leads to explicit expressions for the matrix $\hat{F}_{all}$.  In the absence of grating terms, i.e. for $h=0$, it simplifies to
\begin{eqnarray}
\label{Fnogr}
& & \hat{F}_{h=0}= - \frac{1}{(1+s)^2}  \left(
\begin{array}{cccc}
 p & q \\
p & q
\end{array}
\right) \nonumber
\end{eqnarray}\\
 which leads to $F_{sym} = - \frac{p}{(1+s)^2}$, while $G=1$ as expected, implying a zero eigenvalue for $\hat{F}_{h=0}$, and hence $\psi_1 = \theta$. The MM model gives identical results for $h=0$. \\

 With all  grating terms included, i.e. for $h=1$, we obtain

\begin{eqnarray}
\label{Fall}
& & \hat{F}_{all}=  \left(
\begin{array}{cccc}
(1+s)/W^3 - F & -2q/W^3  \\
-2p/W^3  &(1+s)/W^3  -F^T
\end{array}
\right)
\end{eqnarray}\\
where $F^T(p,q)= F(q,p)$. For equal intensities $W=\sqrt{1+2s }$ and $\xi =0$. Some calculation then shows that $G$ is approximately $2+2s$ for small $s$. The behavior for larger $s$ is dominated by the fact that $F_{11}$ changes sign at $s= 1+ \sqrt{2}$, as does $G$.\\

Turning now to the SFM problem, we note that in the system and models discussed in \cite{labeyrie14}, the origin of the mirror distance coordinate $d$ was at the centre of the cloud. In the present work it is more natural to set the origin at the cloud exit, and to use a dimensionless coordinate $D$, i.e. the mirror is at distance $D L$ beyond the medium. Evidently $d = L(D+\frac{1}{2})$. $D$ can be negative if the feedback optics involves a telescope. The boundary conditions at the output then become $ g(1) = e^{-2\psi_D} f(1)$, where $\psi_D= D \theta $. Using the analysis presented in the Appendix this, along with $f(0)=0$, leads to  the SFM  threshold condition for perfect mirror reflection ($R=1$) \\
\begin{equation}
\label{2LSfbm}
c_1 c_2 +\left(\frac{\psi_2}{\psi_1}c_D^2+\frac{\psi_1}{\psi_2} s_D^2 \right)s_1 s_2 = c_D s_D\left( \beta_1 s_1 c_2 -\beta_2 s_2 c_1\right).
\end{equation}

Here $c_i=cos\psi_i$; $s_i=sin\psi_i$: $c_D=cos\psi_D$; $s_D=sin\psi_D$, and
$\beta_n = \left(\frac{\psi_n}{\theta}-\frac{\theta}{\psi_n}\right)$.

As a first example, we consider $D=0$, i.e. the mirror is directly at the output.   Since $s_D =0$ for $D=0$, (\ref{2LSfbm}) simplifies to \\
\begin{equation}
\label{2LSfbmD0}
c_1 c_2 +\left(\frac{\psi_2}{\psi_1} \right)s_1 s_2 = 0.
\end{equation}

Now,  it is known (e.g.  \cite{Geddes1994}) that (\ref{2LSslab}) can be written as the product of two factors, $H_1 H_2 =0$. An interesting and important feature of  (\ref{2LSfbmD0}) is that it is identical to the condition $H_2 =0$, but with $\psi_i \to 2\psi_i$. Since this corresponds to doubling the length of the medium, we conclude that the transverse instability threshold conditions for a medium with a lossless feedback mirror at its output corresponds exactly to a threshold condition for a medium of twice the length with balanced counterpropagating inputs. This is consistent with the intuitive idea that the SFM system is somehow the "mirror image" of a CP system. There is a twist, however. $H_1 =0$ does not define a threshold for the SFM system at $D=0$, due to the fact that  (\ref{2LSfbmD0}) is not symmetric under  $1 \to 2$.   Geddes et al \cite{Geddes1994} show, using the parity symmetry of the symmetrically pumped CP system, that  $H_1 =0$ and $H_2 =0$ correspond to perturbation eigenmodes  which are respectively odd and even, i.e. $f=-g$ and $f=g$ respectively at the centre of the medium. Only the latter corresponds to the SFM boundary condition, and hence only the even-mode instabilities of the CP system correspond to SFM instabilities. This breaking of parity (and hence $1 \leftrightarrow 2$) symmetry explains why we had to be  careful in defining $\psi_{1,2}$, so as to align them with the Kerr definitions.

%
  \begin{figure}
 \includegraphics[scale=0.9,width=\columnwidth,trim=0 130 0 120,clip]{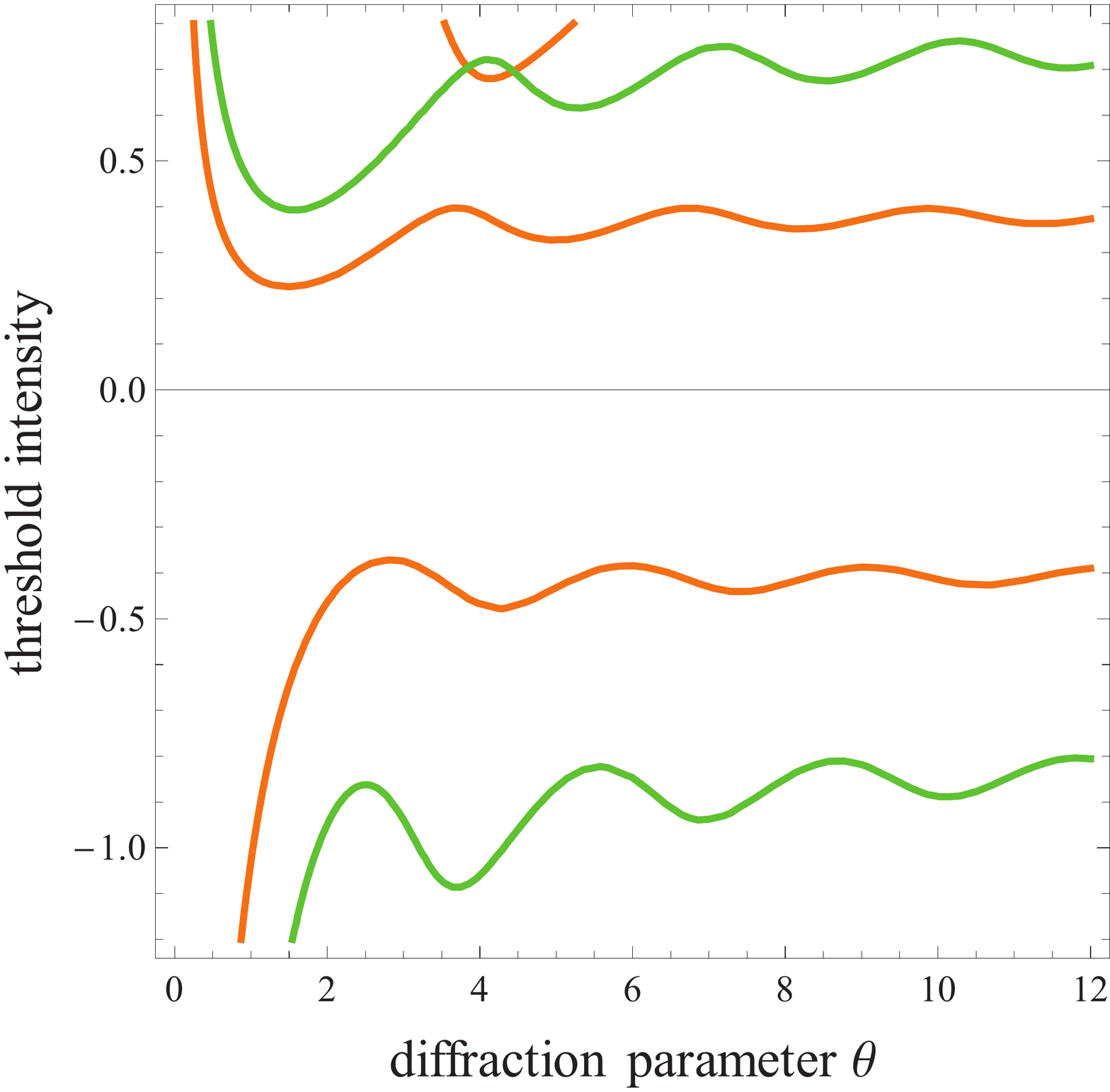}
\caption{
Threshold intensity (in units of $\alpha_l L \Delta p/2$) vs diffraction parameter $\theta =Q^2L/2k$, calculated from (\ref{2LSfbm}) for a Kerr medium  described by $\hat{F}_{kerr}$, with $h=0$ (green) and $h=1$ (orange). Positive and negative intensity values, respectively, correspond to self-focusing and self-defocusing Kerr media. There is a feedback mirror placed directly at the end of the medium ($D=0$, $R=1$).}

 \label{fig:KerrG2D0}
 \end{figure}

Fig.~\ref{fig:KerrG2D0}  shows threshold curves for a Kerr medium with a lossless feedback mirror at its output, calculated from (\ref{2LSfbmD0}) using $D=0$ and $G=1+h=2$. The SFM threshold curves are identical to one of the two intertwined curves found in the CP problem, see, e.g.,  Fig.~2 of \cite{Geddes1994}, allowing for the factor of two in length $L$ needed to align the SFM and SP problems. The intensity unit used in Fig.~\ref{fig:KerrG2D0},  $\alpha_l L \Delta p/2$, is the single-beam, single-pass self-phase-shift of the forward field, as used in   \cite{Geddes1994}, so the twofold reduction in thresholds compared with Fig.~2 of \cite{Geddes1994} is a real advantage of the SFM configuration over the CP one.

\begin{figure}
 \includegraphics[width=\columnwidth,trim=0 245 0 250,clip]{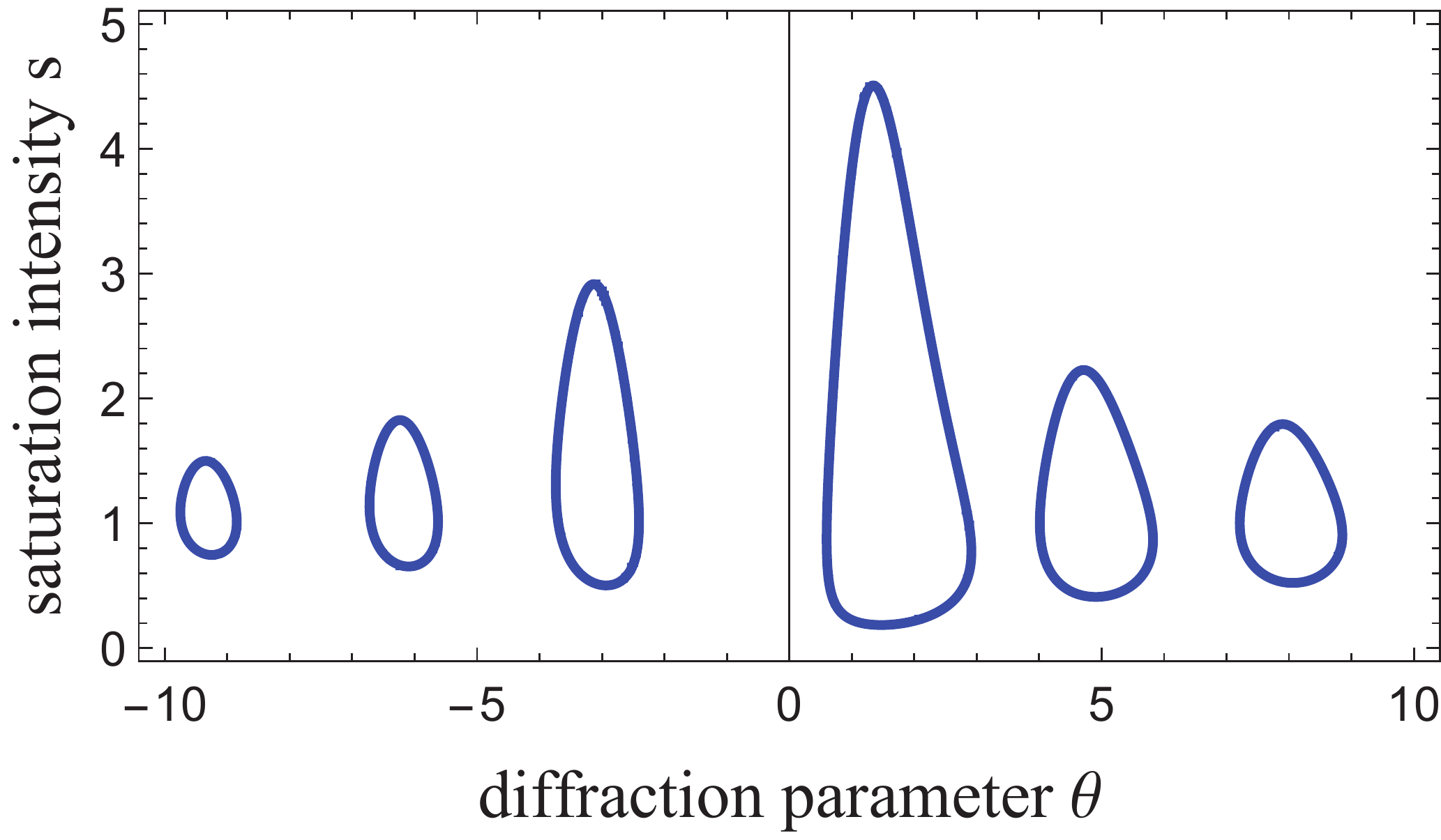}
\caption{
Threshold saturation intensity $s$ vs diffraction parameter $\theta =Q^2L/2k$, calculated from (\ref{2LSfbm}) for a two-level medium  described by $\hat{F}_{All}$, with $h=1$. There is a feedback mirror placed directly at the end of the medium ($D=0$ case). The quasi-Kerr coefficient $|\alpha_l L \Delta|$ is 8, with negative and positive $\theta$ corresponding to red and blue detuning respectively at diffraction parameter $|\theta|$.}

 \label{fig:2LSd0}
 \end{figure}

Turning now to the saturable two-level case, Fig. \ref{fig:2LSd0} shows threshold curves for a two-level medium with a lossless feedback mirror at its output, calculated from (\ref{2LSfbmD0}) using the response function $\hat{F}_{All}$. We have taken advantage of the invariance of the $\psi_i$, and hence of  (\ref{2LSfbmD0}), under simultaneous sign changes of $\theta$ and $\Delta$ to combine red and blue detuning cases in a single graph, with negative $\theta$ corresponding to red detuning. The effect of saturation is clearly seen in the presence of upper, as well as lower, thresholds. The cases shown ($|\alpha_l L \Delta |= 8$) are fairly close to the minimum quasi-Kerr phase shift to allow a transverse instability, so that the unstable domains are closed curves forming distinct bands of unstable transverse wave vectors $Q$. These bands are located in fairly close correspondence to local threshold minima in the  Kerr case (Fig. \ref{fig:KerrG2D0}), with red and blue detuning corresponding to self-defocusing and self-focusing Kerr cases respectively. Hence one  effect of the $f=g$ mode constraint is that the unstable bands for red detuning (left panel) are complementary to (and generally have larger $Q$ than) those for blue detuning (right panel). 

\begin{figure}
 \includegraphics[width=\columnwidth,trim=0 245 0 250,clip]{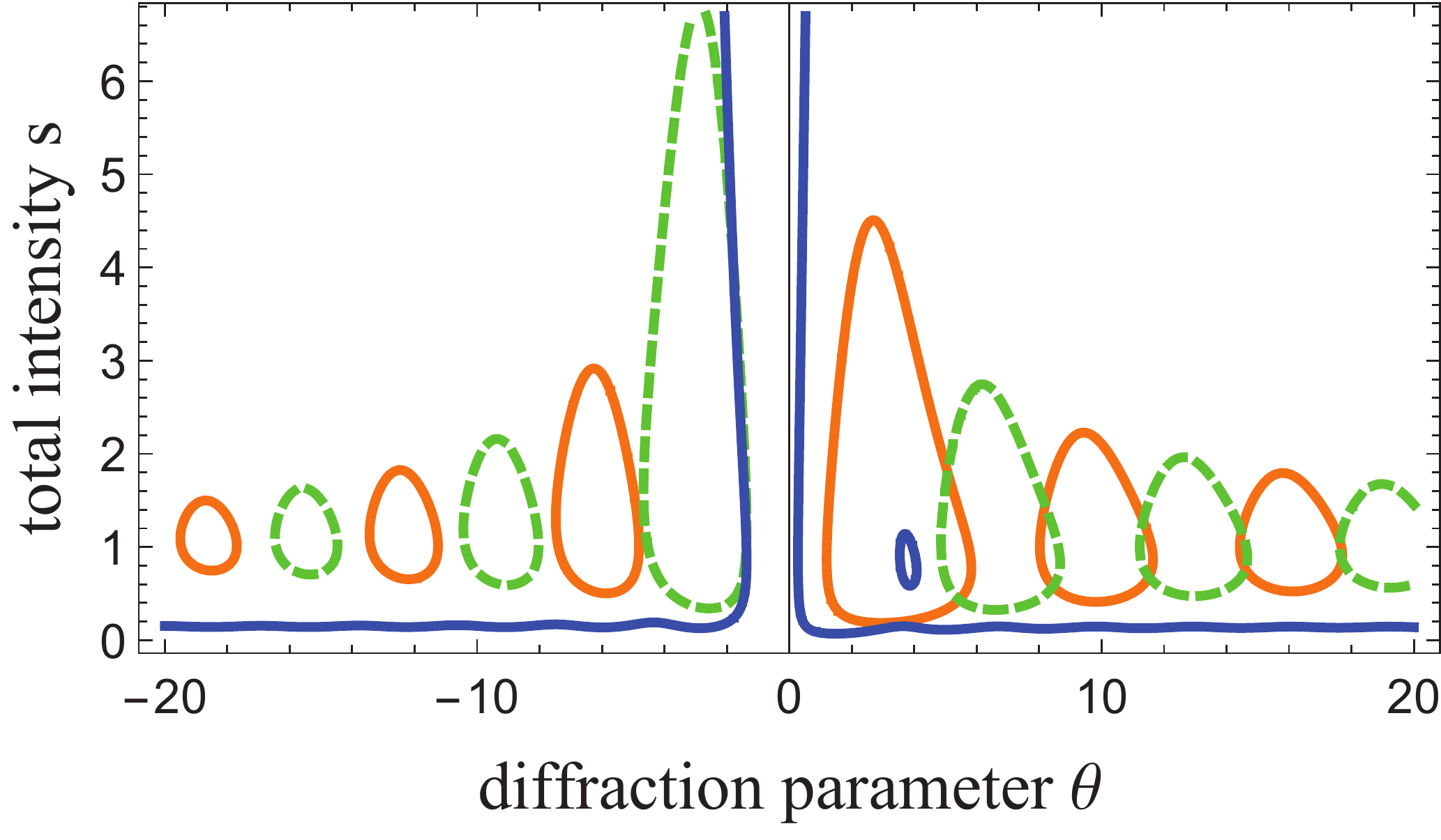}
\caption{
Comparison, for the same two-level medium,  between counterpropagation (CP) and single feedback mirror (SFM) pattern thresholds. Parameters as for Fig.  \ref{fig:2LSd0} except $|\alpha_l L \Delta| =16$. Therefore the closed loops (solid)  marking the even-mode CP instability are identical to those for the SFM instability in Fig.  \ref{fig:2LSd0}, consistent with the scaling relationship discussed in the text. Note the change of  $\theta$ scale from Fig.  \ref{fig:2LSd0}. The odd-mode CP threshold loops (dashed) are absent in Fig.  \ref{fig:2LSd0}, as expected. The SFM thresholds for this case (open curves) are significantly lower than the CP thresholds, while the upper thresholds are  beyond the plot range for $s$. (The small loop at $(\theta, s) \sim (4,1)$ is an island of SFM stability.)  }

 \label{fig:CP_SFM}
 \end{figure}

The scaling relation between the $D=0$ SFM and CP systems applies also to two-level media. 
Fig. \ref{fig:CP_SFM} illustrates the scaling property by doubling the quasi-Kerr coefficient $\alpha_l L \Delta$, equivalent to doubling $L$, compared to  Fig.  \ref{fig:2LSd0}. The CP thresholds appear as closed loops, rather than intertwined open curves as in a Kerr-medium (cf Fig.~\ref{fig:KerrG2D0}), because saturation implies existence of upper, as well as lower instability thresholds. For blue detuning, the loops do intersect at low $\theta$, reminiscent of the Kerr case, though with upper intersections also. The main point to notice, however, is that the even modes (orange) are identical to the corresponding SFM loops in Fig.  \ref{fig:2LSd0}, while the odd modes (green, dashed) are absent from Fig.  \ref{fig:2LSd0}. (Note the change of scale of $\theta$, necessary to align the two cases, since $\theta$ is $ \sim L$.) 

As well as vividly illustrating the scaling relation between the $D=0$ SFM and CP systems, the fact that the (lowest) SFM threshold (blue) is much lower than the CP one (orange, green), illustrates a major practical advantage of the SFM over the CP configuration in terms of achieving instability. Indeed there is no CP instability for the parameters of Fig.  \ref{fig:2LSd0}. Additionally, of course, the SFM configuration
needs only one laser (or half the power compared to splitting a single laser beam to make the two inputs required in the CP configuration).

The SFM thresholds in Fig. \ref{fig:CP_SFM} are rather Kerr-like (though single), with the upper threshold at high enough $s$ to make the curves appear open. The small blue loop is actually an island of SFM stability, which grows as $\alpha_l L \Delta$ is increased. It will occur for CP also, but at still-larger $\alpha_l L \Delta$ and beyond, because of the scaling property.

\begin{figure}
 \includegraphics[width=\columnwidth,trim=0 245 0 250,clip]{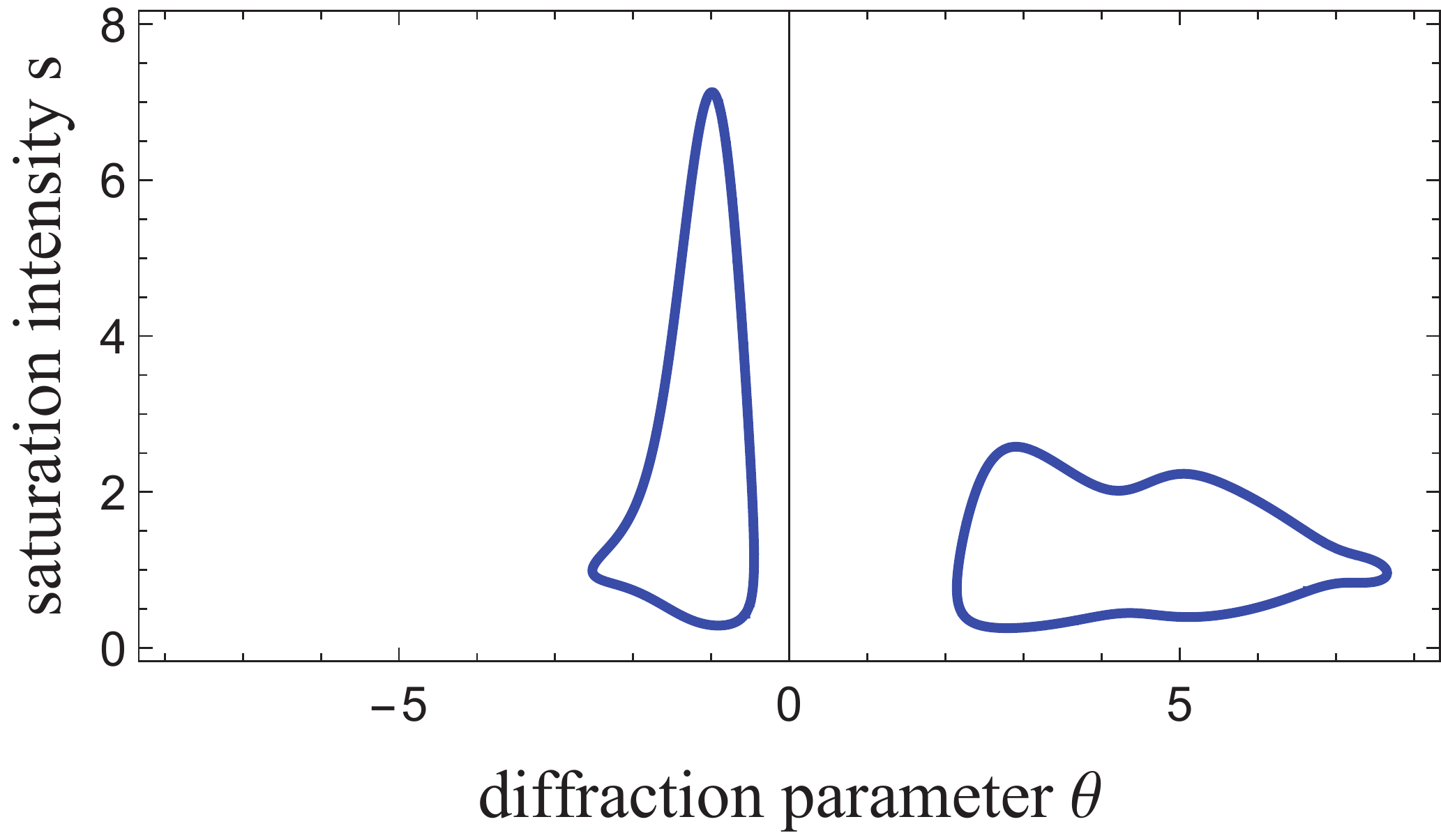}
\caption{
Threshold saturation intensity $s$ vs diffraction parameter $\theta =Q^2L/2k$, calculated from (\ref{2LSfbm}) for a two-level medium  described by $\hat{F}_{All}$, with $h=1$. There is a feedback mirror at negative effective distance ($D=-1.3$) from the end of the medium. The quasi-Kerr coefficient $|\alpha_l L \Delta| = 8$, with negative and positive $\theta$ corresponding to red and blue detuning respectively  at diffraction parameter $|\theta|$.}

 \label{fig:2LSdneg177}
 \end{figure}

Fig. \ref{fig:2LSdneg177} shows the effect of finite mirror distance $D$, in this case negative, for blue/red detuning. Because the finite-$D$ formula (\ref{2LSfbm}) is also invariant under simultaneous sign changes of $\theta$ and $\Delta$, we can again use negative $\theta$ to display both blue and  red detuning thresholds on the same graph. The thresholds are somewhat lowered in comparison with $D=0$, and the unstable bands are shifted as well as broadened. Indeed the red detuning now has an unstable band at small $Q$, corresponding to a small-angle scattering cone in the far field. This sensitivity to mirror distance can be interpreted as a phase-matching effect: the external phase shift $\psi_D$ provides an extra flexibility in comparison with the CP problem, enabling instability in cases where the internal nonlinear and diffractive phases are ill-matched.

%
%

%
%

\begin{figure}
 \includegraphics[scale=0.9,width=\columnwidth,trim=0 130 0 120,clip]{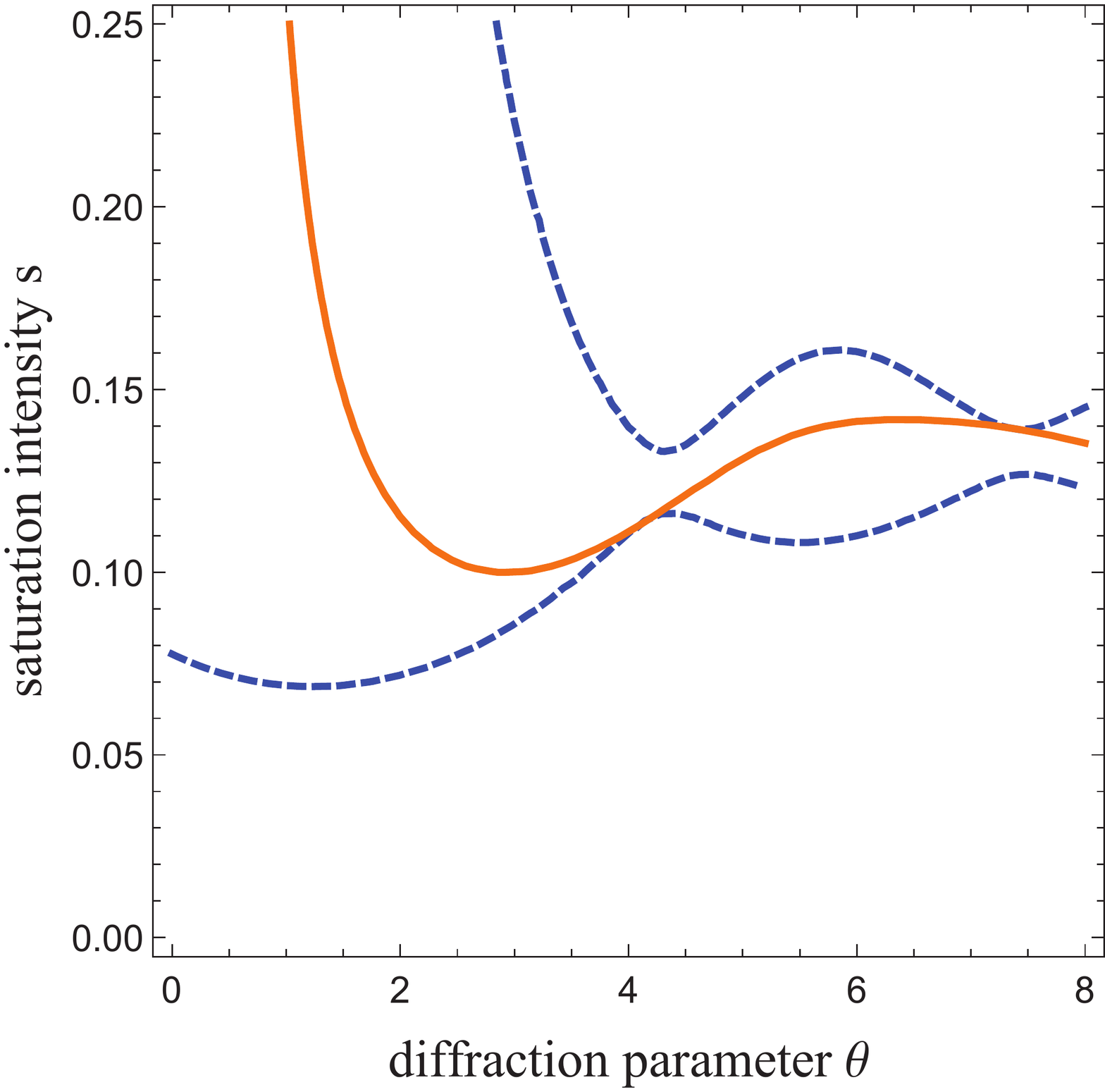}
\caption{ Saturation intensity $s$ vs diffraction parameter $\theta =Q^2L/2k$. Blue curves: Envelope curves  calculated from (\ref{env}) for a two-level medium  described by $\hat{F}_{All}$, with $h=1$. The quasi-Kerr coefficient $|\alpha_l L \Delta| =16$, corresponding to blue detuning. Orange curve: Threshold curve with a feedback mirror at negative effective distance ($D=-0.5$, i.e. at the centre of the medium), which touches the envelope curves. }

 \label{fig:2LSenv}
 \end{figure}

\begin{figure}
 \includegraphics[scale=0.9,width=\columnwidth,trim=0 115 0 110,clip]{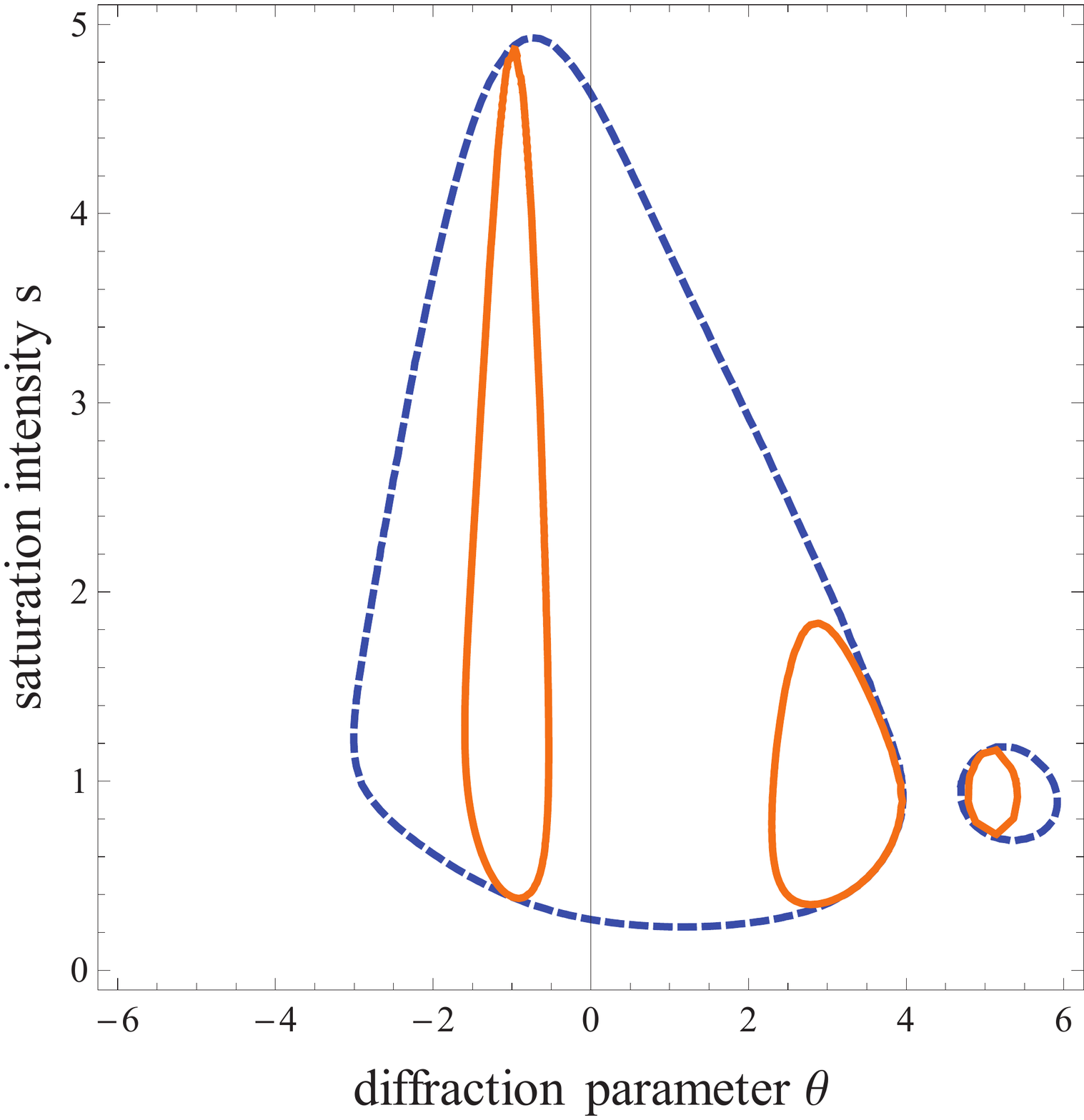}
\caption{ Saturation intensity $s$ vs diffraction parameter $\theta =Q^2L/2k$. Blue curves (dashed): Envelope curves  calculated from (\ref{env}) for a two-level medium  described by $\hat{F}_{All}$, with $h=1$. Quasi-Kerr coefficient $|\alpha_l L \Delta| =7.1$. Orange curves: Threshold curves with a feedback mirror at negative effective distance ($D=-1.3$) from the end of the medium, which touches the envelope curves.}

 \label{fig:2LSenv_extra}
 \end{figure}

The presence of mirror distance as an additional parameter in the SFM formula (\ref{2LSfbm}) as compared to the CP formula (\ref{2LSslab}) makes it harder to see what is going on. Especially at small $|D|$, the transverse wavelength with lowest threshold varies strongly with mirror distance. The threshold intensity, however, varies much less, and we now show that the threshold curve (intensity vs diffraction parameter $\theta$) is bounded below by an envelope curve. Indeed there are a set of upper and lower envelope curves, which can be calculated analytically from (\ref{2LSfbm}).

In deriving (\ref{2LSfbm}), we naturally assumed that the feedback phase functions $c_D, s_D$ are real. If the intensity is just below a threshold minimum, however, we can still find a solution with complex $\psi_D$, which corresponds physically to introducing some gain into the feedback loop. Just above the minimum, there are generally two adjacent values of $\theta$ which solve (\ref{2LSfbm}). Since the threshold curves oscillate, with maxima as well as minima, we note that two real roots just below a maximum become no roots just above, again with complex roots corresponding to feedback gain.

We can quantify this scenario by observing that (\ref{2LSfbm}) can be turned into a quadratic equation in $tan \psi_D$. Vanishing discriminant for this equation corresponds to the transition between complex and real $\psi_D$.  This results in the following equation:

\begin{equation}
\label{env}
4(c_1 c_2 +\frac{\psi_1}{\psi_2} s_1 s_2)(c_1 c_2 +\frac{\psi_2}{\psi_1} s_1 s_2) = ( \beta_1 s_1 c_2 -\beta_2 s_2 c_1)^2.
\end{equation}

As a first example,  Fig.~\ref{fig:2LSenv} illustrates the envelope curves for the all-grating quasi-Kerr model, together with a part of the threshold curve for $D= -0.5$, which indeed touches both curves. The system parameters here are similar to those in Fig. \ref{fig:CP_SFM}, and thus enable a more detailed view of the shape of the SFM threshold curve, as well as the very low $s$ values at which thrshold can be reached.   Contact with the envelope is not necessarily at the extrema of the threshold curve, but for all $D$ values, and all cases, considered the threshold curves are bounded by, and tangent to, the envelope curves given by (\ref{env}). In particular, the absolute minimum of the envelope, approximately at $\Theta=1.5$, $s=0.08$ in  Fig.~\ref{fig:2LSenv}, defines the minimum attainable threshold for  $|\alpha_l L \Delta| = 16$ with other medium parameters fixed. \\

Whereas the threshold curves either asymptote to, or are distinct from the axis $\theta =0$, the envelope curve in Fig.~\ref{fig:2LSenv} seems to approach the axis at finite $s$. This is confirmed and eloborated in Fig. \ref {fig:2LSenv_extra}, where the trick of plotting also for $\theta < 0$ nicely exhibits the finiteness of the intercept of the envelope, as well as continuity across the  $\theta =0$ axis between red and blue detuning. Our interpretation is that the envelope intercept corresponds to $L \to 0$, i.e. the "thin medium" limit, which we will discuss further shortly. First, however, we note that the thresholds for $D=-1.3$ are neatly tangential to the envelope curves, including the small envelope loop at $\theta \sim 5$. Since there is no corresponding loop for red detuning, we can conclude that only a single band of patterns can be found, at any $D$, for red detuning. Another important feature of Fig. \ref {fig:2LSenv_extra} is that the finite slope of the envelope at the axis means that one or other of the detunings has its absolute minimum threshold at finite $\theta$, whereas for the other the threshold decreases as $D$ is increased, with minimum threshold being found in the thin-medium limit.

\begin{figure}

 \includegraphics[scale=0.9,width=\columnwidth,trim=0 130 0 120,clip]{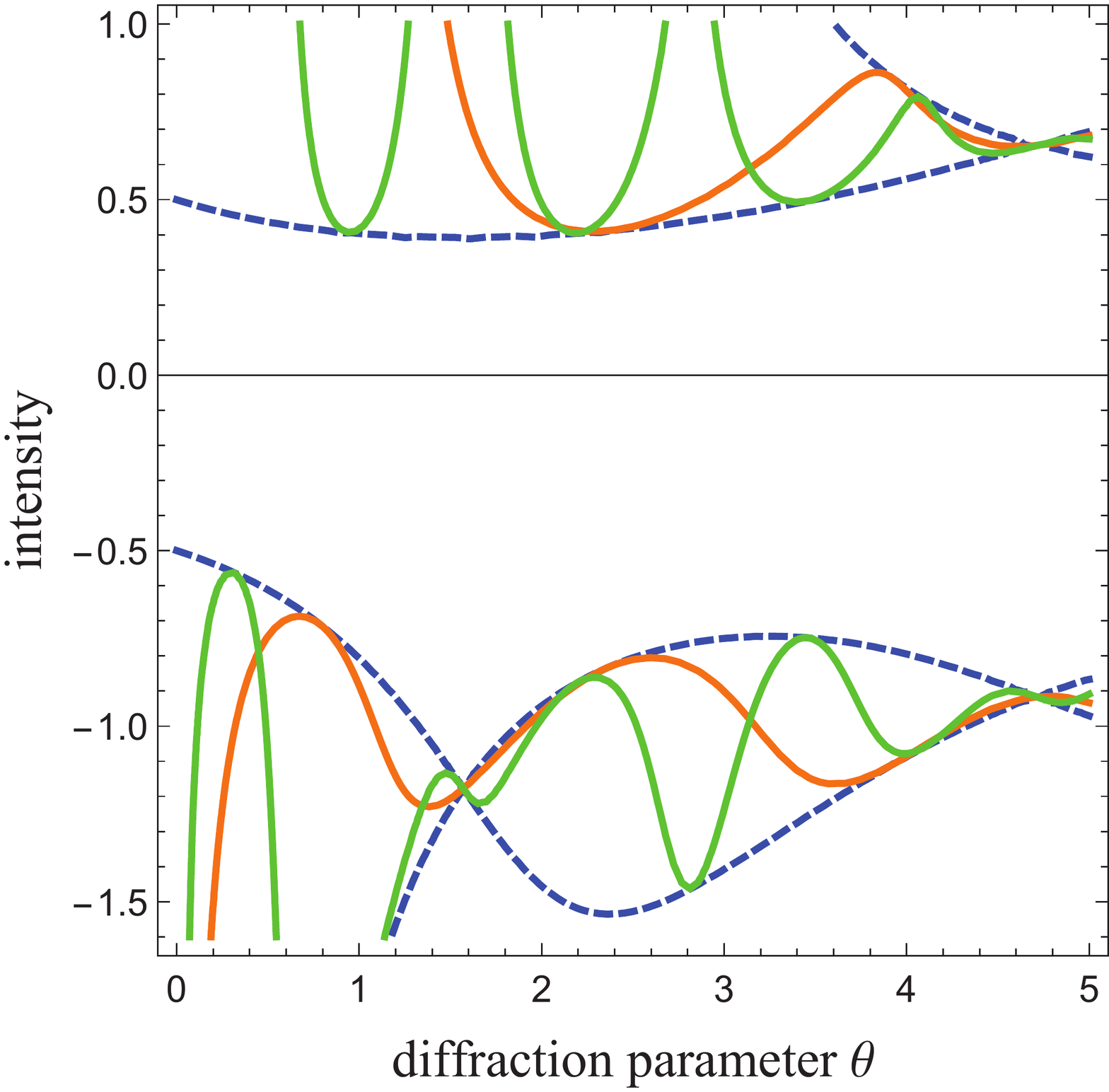}
\caption{ Threshold intensity (in units of $\alpha_l L \Delta p/2$)  vs diffraction parameter $\theta =Q^2L/2k$. Blue curves: Envelope curves calculated from (\ref{env}) for a Kerr medium with $h=0$, i.e. $G=1$. Positive and negative intensity values, respectively, correspond to self-focusing and self-defocusing Kerr media. Also threshold curves with a feedback mirror at negative effective distance from the end of the medium. Orange curves: $D=-1.5$. Green curves: $D=-3.0$. In both cases the threshold curves touch the envelope curves, and are confined by them.}
 \label{fig:KG1env}
 \end{figure}

Figure~\ref{fig:KG1env} further illustrates how the envelope curves capture the essential behavior of the threshold curves, this time for a Kerr medium with no grating term ($h=0$). Here two distances ($D=-1.5,-3.0$) are shown, and we begin to see how the faster oscillations of the threshold for larger mirror distances allow a better exploration of the envelope, and thus potentially lower thresholds. For the self-focusing case, where the envelope has a minimum at finite $\theta$, we can see, for $D=3$, the transition of the lowest threshold  from the lowest-Q to the second-lowest-Q band. Assuming that the dominant pattern is determined by the lowest threshold, we would expect a sudden drop in the observed pattern perod as $D$ is increased. This phenomenon is indeed observed (see Fig.~\ref{pol_Talbotscan}  below for an example). Conversely, for self-defocusing the lowest  threshold always decreases as $D$ is increased, so that the patterns with lowest threshold are found at large mirror distances, and have large spatial scales, with pattern wavelength scaling like $\sqrt{d/k}$, as is well known from thin-medium theory \cite{firth90a}. In contrast, CP thresholds for $G=1$ defocusing Kerr media decrease with increasing $Q$, see, e.g.  \cite{Geddes1994}. The same is true, of course, for the SFM with $D=0$, as shown in Fig.~\ref{fig:KerrG2D0}. This finite-$D$ advantage can be attributed to the ability of the feedback phase to compensate for both the diffractive and nonlinear phase shifts in the medium, which have the same sign for defocusing, and thus cannot cancel each other as they can for self-focusing. This no-grating Kerr case is also interesting in that the envelope curves cross, and hence the threshold curves must thread through the intersection (Fig.~\ref{fig:KG1env}). It follows that  the threshold is actually independent of mirror distance at these crossings. Note that the  threshold will normally be lower at a different diffraction parameter (as occurs in Fig.~\ref{fig:KG1env}),  and observing the phenomenon would require isolating the  specific wavenumber by Fourier filtering in the feedback loop \cite{pesch03}. 

\begin{figure}
 \includegraphics[scale=0.9,width=\columnwidth,trim=0 130 0 120,clip]{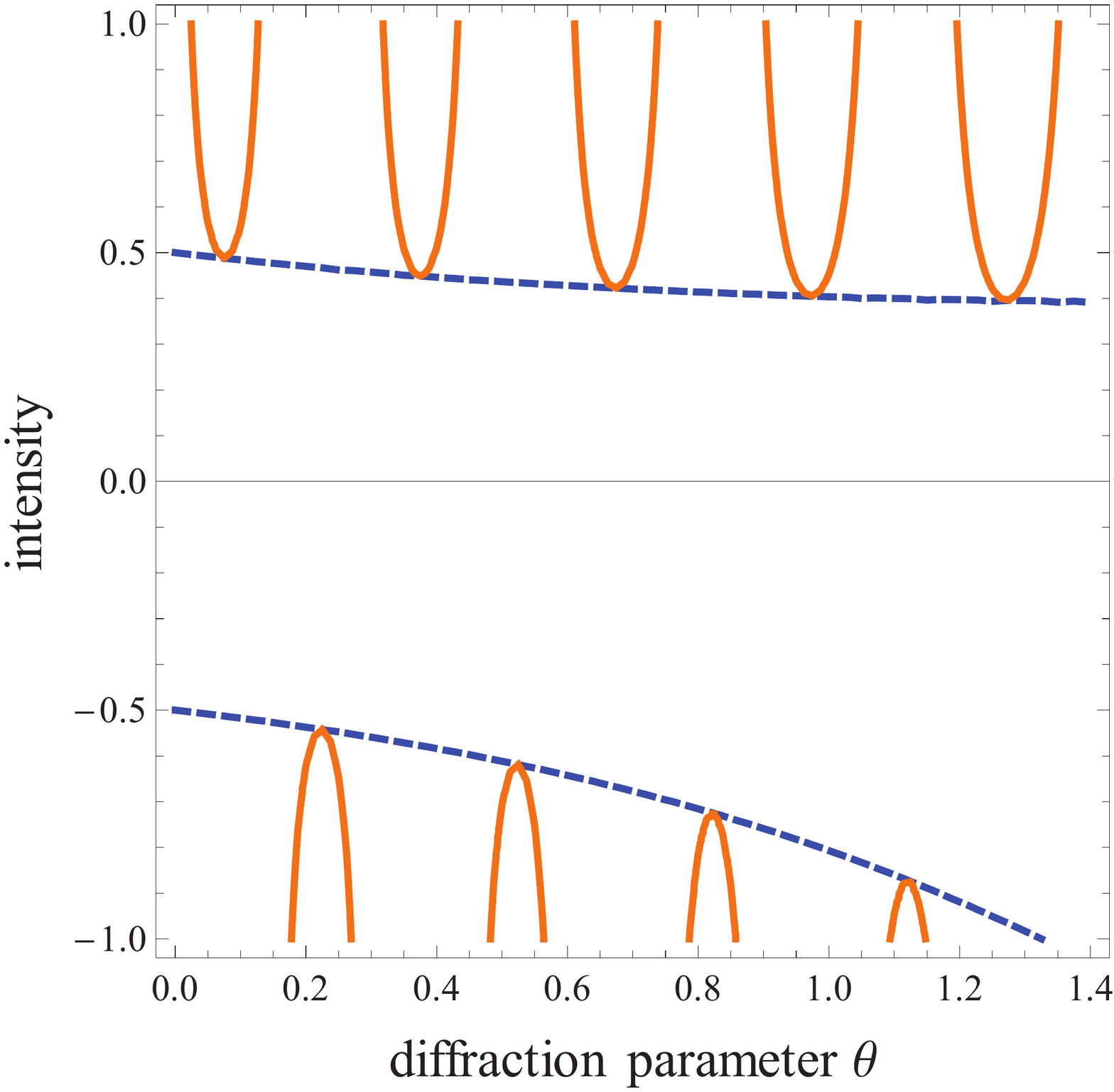}
\caption{
Threshold intensity (in units of $\alpha_l L \Delta p/2$) vs diffraction parameter $\theta =Q^2L/2k$. Blue curves: envelope curves calculated from (\ref{2LSfbm}) for a Kerr medium  described by $\hat{F}_{kerr}$, with $h=0$, i.e.\ $G=1$. Positive and negative intensity values, respectively, correspond to self-focusing and self-defocusing Kerr media. Orange curves: $D=10$. The feedback mirror is quite far from the medium, which is thus a quite-thin slice. Note that the mode thresholds are not degenerate, as they are in simple thin-slice SFM models \cite{firth90a}.}

 \label{fig:KerrG1Dlarge}
 \end{figure}

The finite limit for small diffraction, $\theta \to 0$, of the envelope is ($\pm 0.5$) in Fig. \ref{fig:KG1env}, and corresponds exactly to the thin-slice value  \cite{firth90a}, but the finite slope at $\theta=0$ means that the pattern-forming modes are not, in fact, threshold-degenerate when the medium thickness is taken into account. Fig. \ref{fig:KerrG1Dlarge} shows this in more detail for a moderately large mirror distance ($D=10$). For self-focusing the envelope curve falls for increasing diffraction parameter in the range displayed. The minimum is reached at $\theta=\pi/2$. For negative detuning the lowest wavenumber is selected. In both cases, therefore,  the multi-fractal patterns predicted in the thin-slice limit \cite{Huang2005} and dependent on mode-degeneracy are not expected to occur in practice, unless other mechanisms or devices are able to restore degeneracy.

%
%

%

\begin{figure}

 \includegraphics[width=\columnwidth,trim=0 130 0 120,clip]{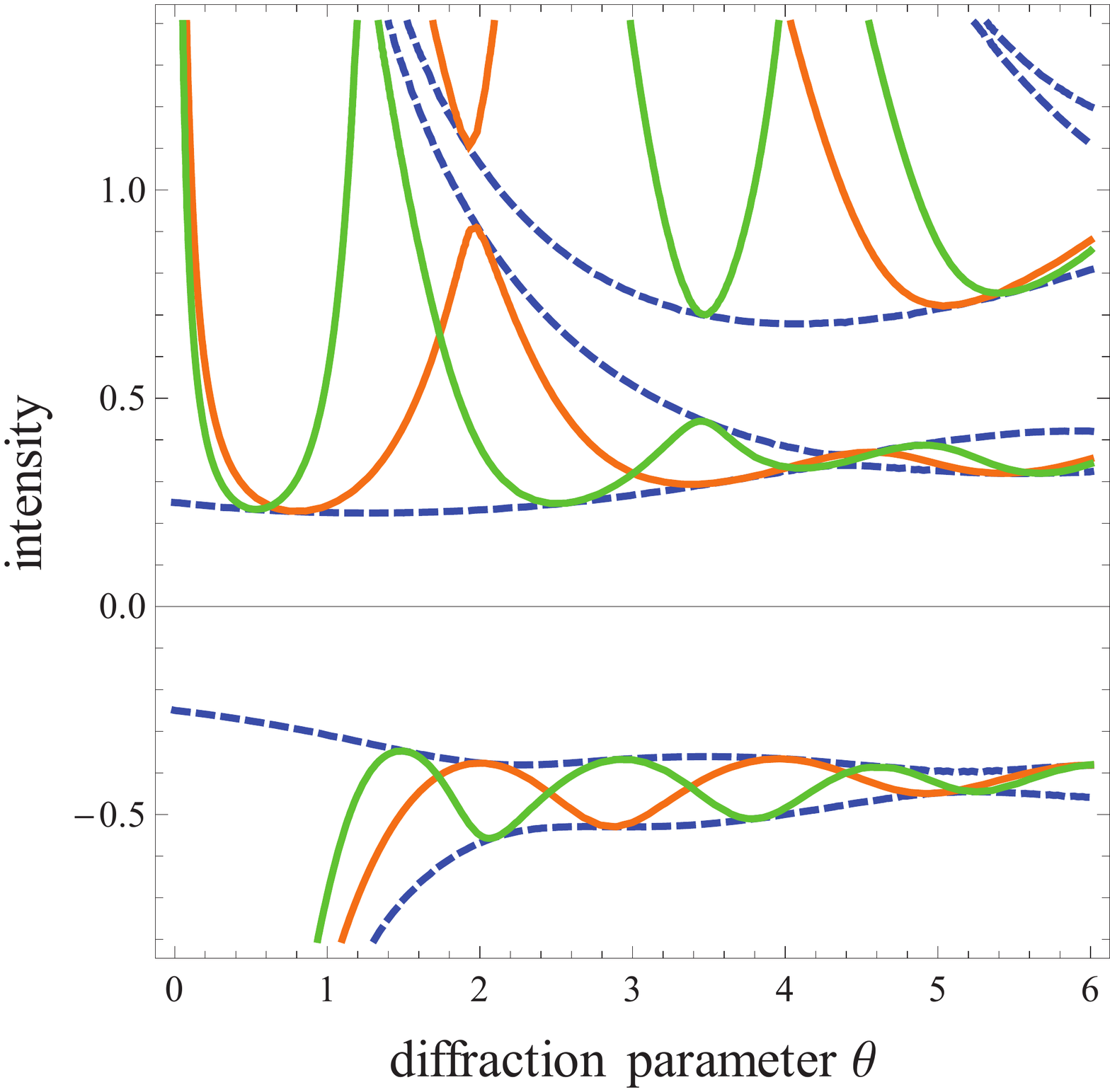}
\caption{ Threshold intensity (in units of $\alpha_l L \Delta p/2$) vs diffraction parameter $\theta =Q^2L/2k$ for a Kerr medium with $h=1$. Negative intensities correspond to negative Kerr, i.e. self-defocusing. Blue curves: Envelope curves  calculated from (\ref{env}). Also threshold curves with a feedback mirror at positive effective distance. Orange curves: $D=0.5$. Green curves: $D= 1.0$. In both cases the threshold curves touch the envelope curves, and are confined by them.}

 \label{fig:KG2env}
 \end{figure}

Further envelope properties are illustrated by the envelope curves for a Kerr medium with grating  (Fig. \ref{fig:KG2env}), this time plotted along with threshold curves for positive mirror distances. In this case higher-order modes are visible, but the corresponding envelope curves again confine the corresponding threshold curves. Here the envelopes of the lowest order modes do not actually cross, though there are still values of $\theta$ for which the threshold is almost distance independent. Again the small-diffraction limit corresponds to the standard thin-slice threshold, but this limit is approached with finite slope.

%
%

\begin{figure}
 \includegraphics[width=\columnwidth,trim=0 130 0 120,clip]{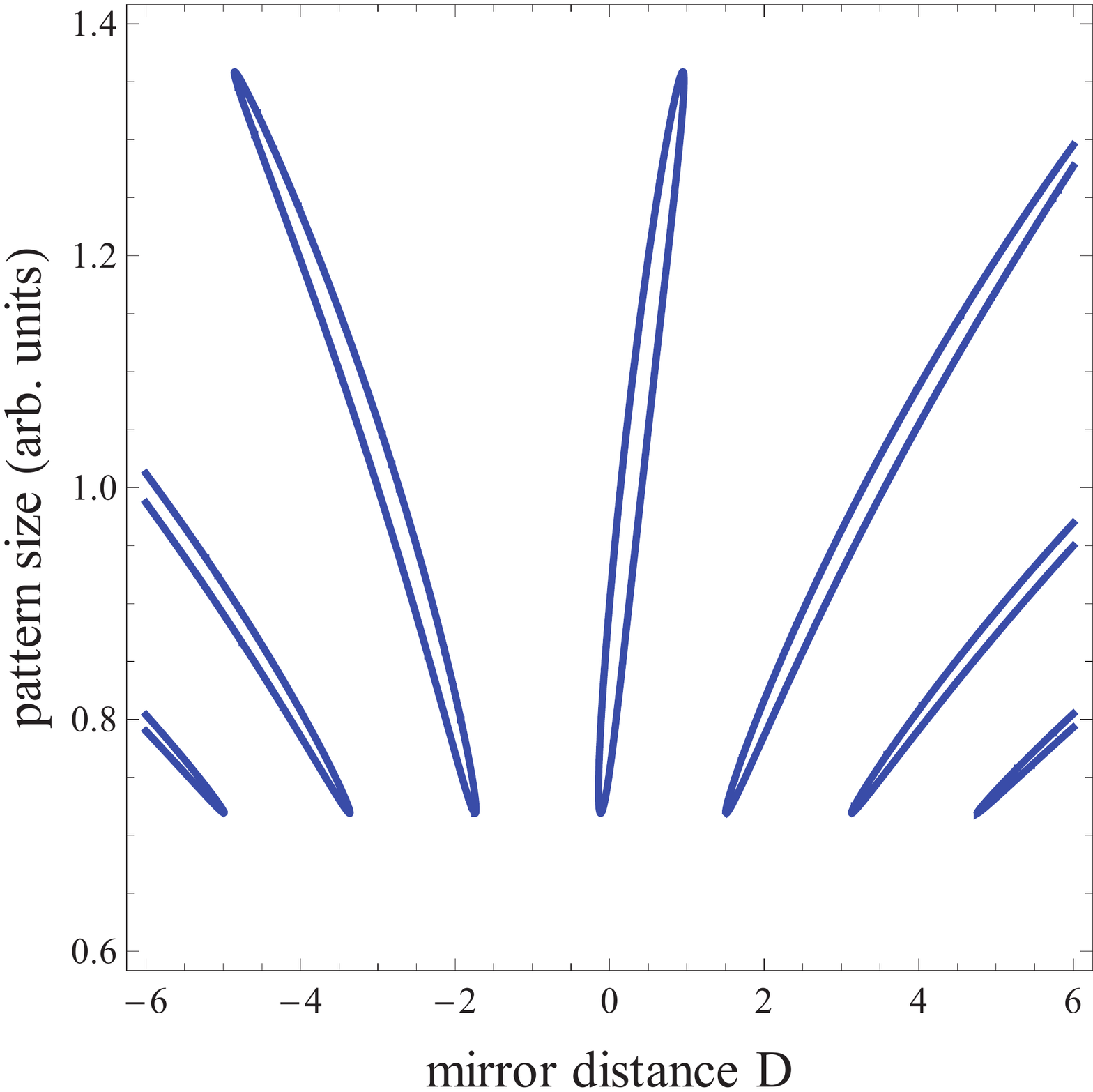}
\caption{ Pattern period (arb.\ units)  vs mirror distance $D$ at fixed intensity $s=0.085$. Threshold curves  calculated from (\ref{2LSfbm}) for a two-level medium  described by $\hat{F}_{All}$, with $h=1$. The quasi-Kerr coefficient $\alpha_l L \Delta =13.94$, corresponding to blue detuning. For optical density 210 \cite{Camara2015}, this corresponds to detuning $\Delta = 2\delta/\Gamma = 15$.}
 \label{fig:2LSsize}
 \end{figure}

The above figures demonstrate how the threshold extrema move vs $\theta$ as mirror distance $D$  is varied. An interesting and relevant way to examine this is to plot pattern scale ( $ \sim 1/\sqrt{\theta}$) vs $D$ for fixed intensity. This is demonstrated in Fig.~\ref{fig:2LSsize}, where the parameters are chosen to match those of  \cite{Camara2015}, and the intensity $s=0.085$ is just above the minimum threshold, so that the unstable regions appear as long narrow islands. The "fan" shape of the island group is due to the Talbot effect: the threshold values satisfying (\ref{2LSfbm}) are evidently periodic in $\psi_D = D \theta$, which means that at fixed $\theta$ (size) and intensity, threshold values are periodic in $D$. This is particularly clear at the bottom of the fan in Fig. \ref{fig:2LSsize}, where the tips of the islands are equally-spaced in $D$. The Talbot periodicity is inversely proportional to $\theta$, which is why the islands fan out as the pattern scale increases (i.e. as $\theta$ decreases).

\begin{figure}
 \includegraphics[width=\columnwidth,trim=0 60 0 20,clip]{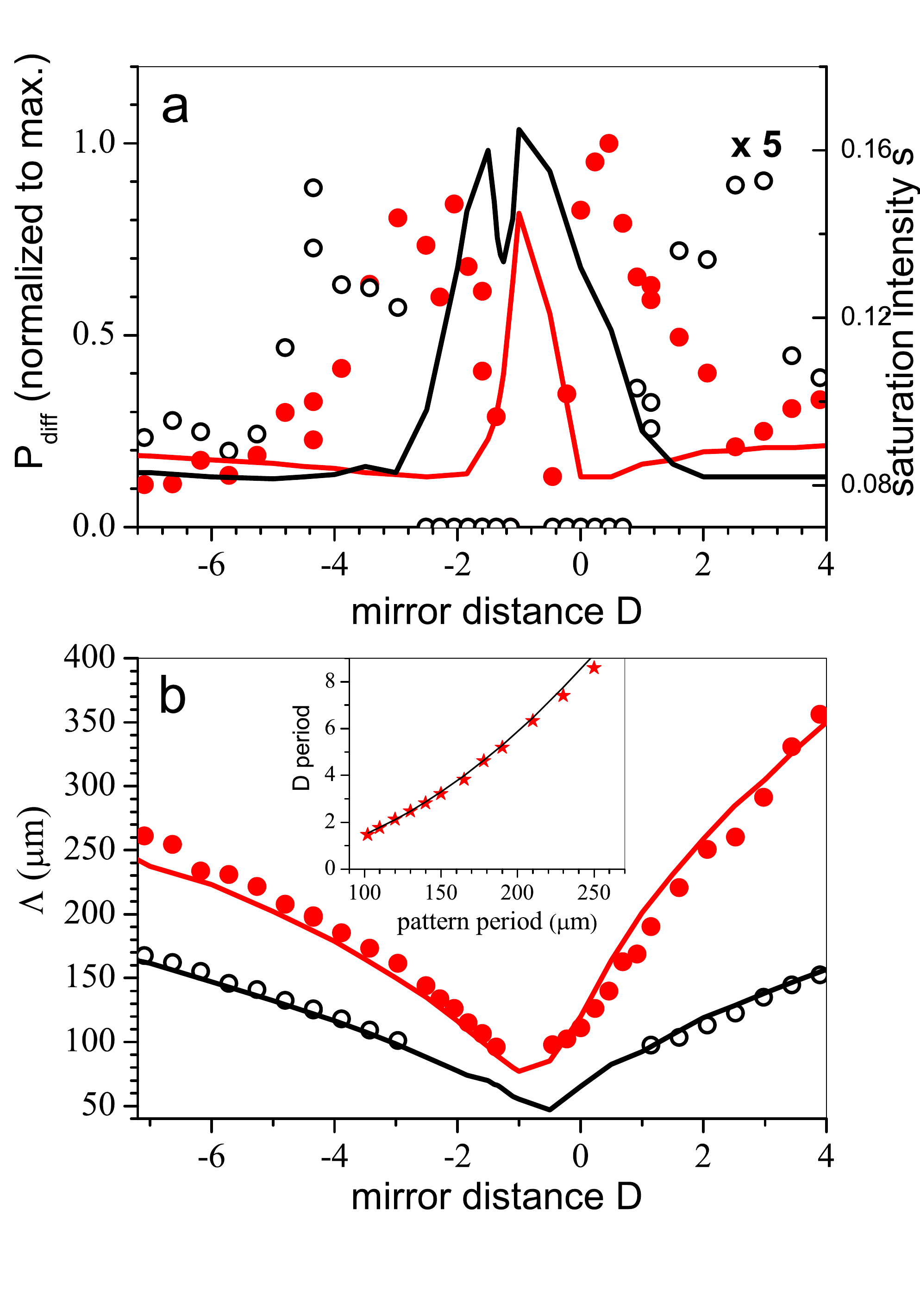}
\caption{(Color online) a) Diffracted power (experiment, left axis) and predicted threshold saturation intensity (theory, right axis) vs scaled mirror distance $D$. The cloud thickness is $L=9$ mm. b) Pattern period  $\Lambda$ vs mirror distance. In physical units, the x-axis corresponds to -60~mm to +40~mm measured from the center of the cloud. Parameters: blue detuning, $\Delta = 15$, see \cite{Camara2015}. The diffracted power is normalized to its maximal value. Red solid dots: experimental data for first Talbot balloon (lowest wavenumber), black circles: experimental data for second Talbot balloon (next highest wavenumber excited, in a) enhanced by factor of 5). The red and black curves are the corresponding theoretical predictions and are calculated from (\ref{2LSfbm}) using the all-grating two-level model.
 Inset: The measured $D$ period as a function of the pattern size (stars), together with the Talbot effect prediction (line). }
 \label{Talbotfan}
 \end{figure}

\begin{figure}
 \includegraphics[scale=0.9,width=\columnwidth,trim=10 80 5 75,clip]{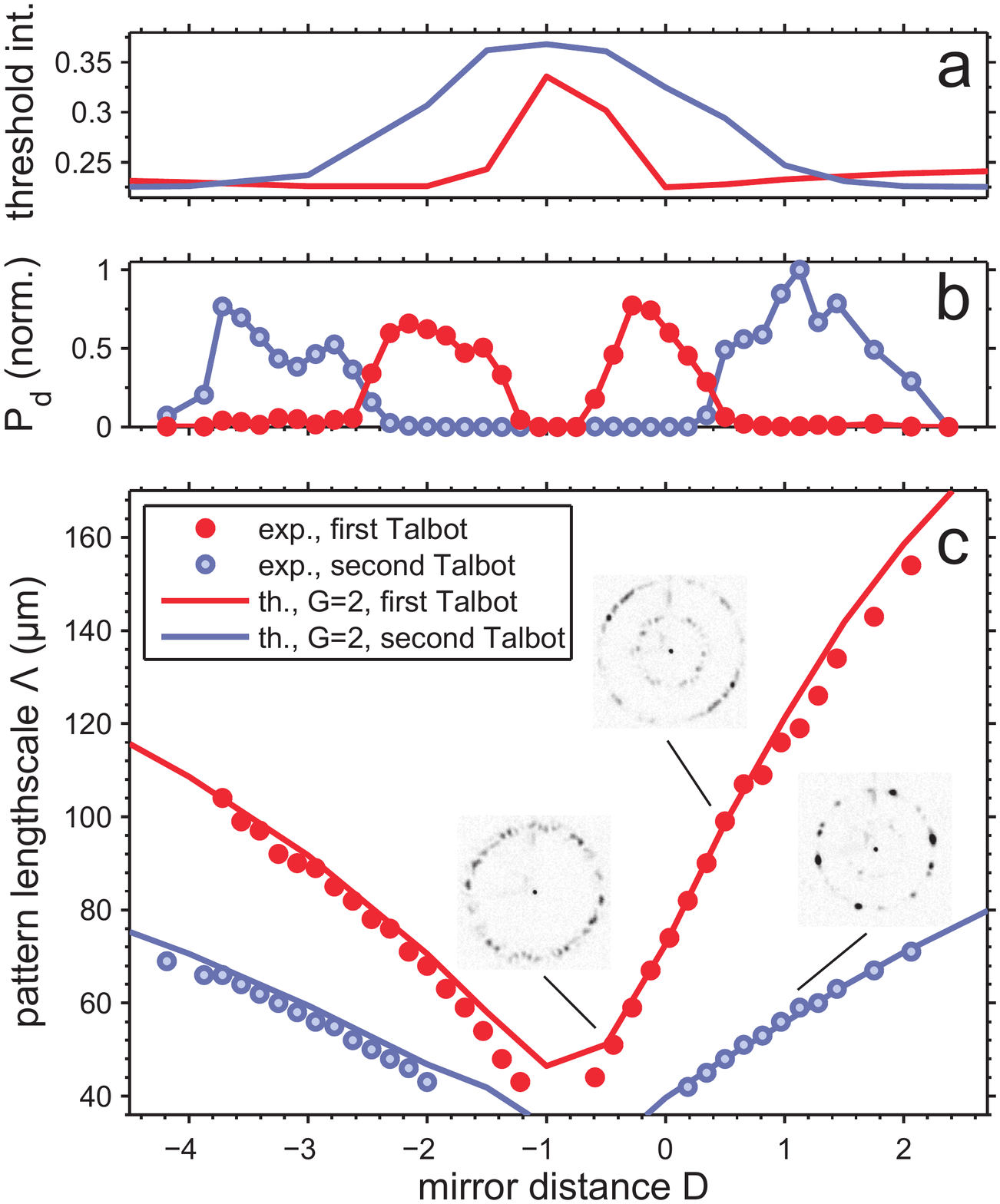}
\caption{ (Color online) a) Predicted threshold, b) experimentally observed diffracted power (normalized to its maximal value) and c) pattern period vs mirror distance $D$. In unscaled parameters, the x-axis corresponds to -12.8~mm to +10.2~mm measured from cloud center. Parameters: effective medium length is $L=3.2$ mm, beam intensity $I=18$ mW/cm$^2$ and detuning $\Delta = -14$. Red solid dots: experimental data for first Talbot balloon (lowest wavenumber), blue circles: experimental data for second Talbot balloon (next highest wavenumber excited). The red and blue curves are the corresponding theoretical predictions and are calculated for a self-focusing Kerr medium with $h=1$ described by $\hat{F}_{Kerr}$. The insets show far field patterns obtained at the mirror positions indicated illustrating the length scale competition. }
\label{pol_Talbotscan}
\end{figure}

Such "Talbot fans" are readily observed experimentally. The fan reported in \cite{Camara2015} is shown in Fig.~\ref{Talbotfan}, where the experimental data fit well to threshold data from  (\ref{2LSfbm}) using our two-level all-grating model based on $\hat{F}_{all}$. Fig.~\ref{Talbotfan}b plots the pattern period against mirror distance. Around $D\approx 0$ the lengthscale with the smallest wavenumber (largest period) is selected. At higher $|D|$, two lengthscales are found in the pattern. Both are in good agreement with the prediction from the theory. The inset shows excellent agreement between the measured and calculated $D$-periodicities. In the earlier optomechanical patterns paper \cite{labeyrie14}, there is a more limited fan, to which threshold data from (\ref{2LSfbm}) are fitted using a Kerr model (h=0, because the slow time scale allows atomic motion to wash out the longitudinal grating).

Fig.~\ref{Talbotfan}a plots the power diffracted into the first and second unstable wavenumber obtained by integrating the measured far field intensity distributions over an annulus with the respective radius. We did not measure thresholds, but to a first approximation one can argue that the diffracted power increases with increasing distance to threshold and hence the measured data can be interpreted as indicators of inverted threshold curves. We compare them with the threshold curves obtained from the all grating quasi-Kerr model as the detuning is reasonably large and absorption not very important. As indicated in the discussion of  Fig.~\ref{Talbotfan}a, around $D\approx 0$, only the lowest wavenumber (i.e. the one from the first Talbot balloon) is excited. For a mirror within the medium ($D=-1 \dots 0$), the diffracted power is low and the predicted thresholds are high. For increasing $|D|$ threshold are predicted to fall dramatically and indeed well developed patterns, indicated by high diffracted power, are observed. For further increasing $|D|$ the theory predicts that the second Talbot balloon at higher wavenumber has the lowest threshold. Indeed excitation of this length scale is observed but it does not take over completely in the experimental data.

For a further investigation of the Talbot fan phenomenon we analyze a somewhat different experimental SFM situation in which optical pumping between Zeeman substates, rather than two-level electronic excitation, is the main nonlinearity \cite{grynberg94,scroggie96,leberre95b,aumann97}. Experimental parameters are an effective medium length of $L=3.2$ mm, beam intensity $I=18$ mW/cm$^2$ and detuning $\Delta = -14$. The homogenous solution is not saturated in this case \cite{ackemann01b}, so it is reasonable to compare the data to the length scales and threshold curves obtained from a self-focusing thick medium Kerr theory.

Experimental measurements of diffracted power and pattern lengthscale vs mirror distance are shown in Fig.~\ref{pol_Talbotscan}. It is apparent that the behavior is very similar to the one observed for the electronic 2-level case in Fig.~\ref{Talbotfan}, but there is one crucial difference. For large enough $|D|$ ($D> 0.7$, $D<-2.5$) the length scale from the first Talbot balloon is completely suppressed and the length scale of the second balloon takes over completely. This is in good, although not quantitative, agreement with the thick medium model as discussed earlier in connection with Figure~\ref{fig:KG1env}, though the transition is predicted to occur at somewhat larger $|D|$. Nevertheless, it is an important confirmation of the importance of the diffraction within the medium influencing length scale selection. In view of the fact that the atomic clouds have an approximately Gaussian density distribution and the theory assumes a rectangular distribution, quantitative deviations between theory and experiment are not surprising.

%
%

Figures~\ref{Talbotfan} and \ref{pol_Talbotscan} indicate that a change of mirror distance can drag the pattern period along qualitatively as in a diffractively thin medium but only up to a point. Then the system jumps back to a smaller length scale it seems to prefer, which can be changed again to some extent by changing mirror distance. The origin of this behavior lies in the interaction between the threshold curves and the envelope as discussed before. For increasing $|D|$ the threshold curves move to lower $Q$ and have more wiggles in a certain range of $\theta$ on the envelope curve, which means they can explore more effectively the potentially lowest threshold condition.

\begin{figure}
 \includegraphics[width=\columnwidth,clip]{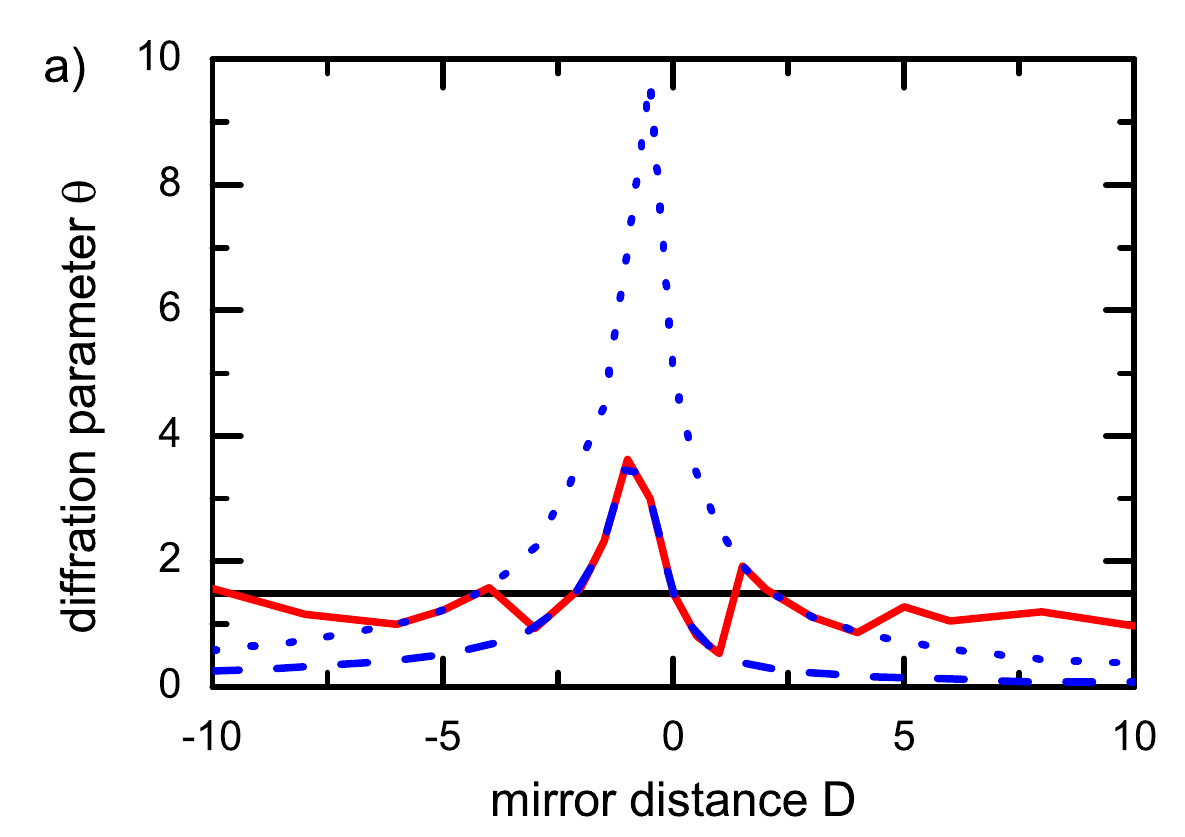}
\includegraphics[width=\columnwidth,clip]{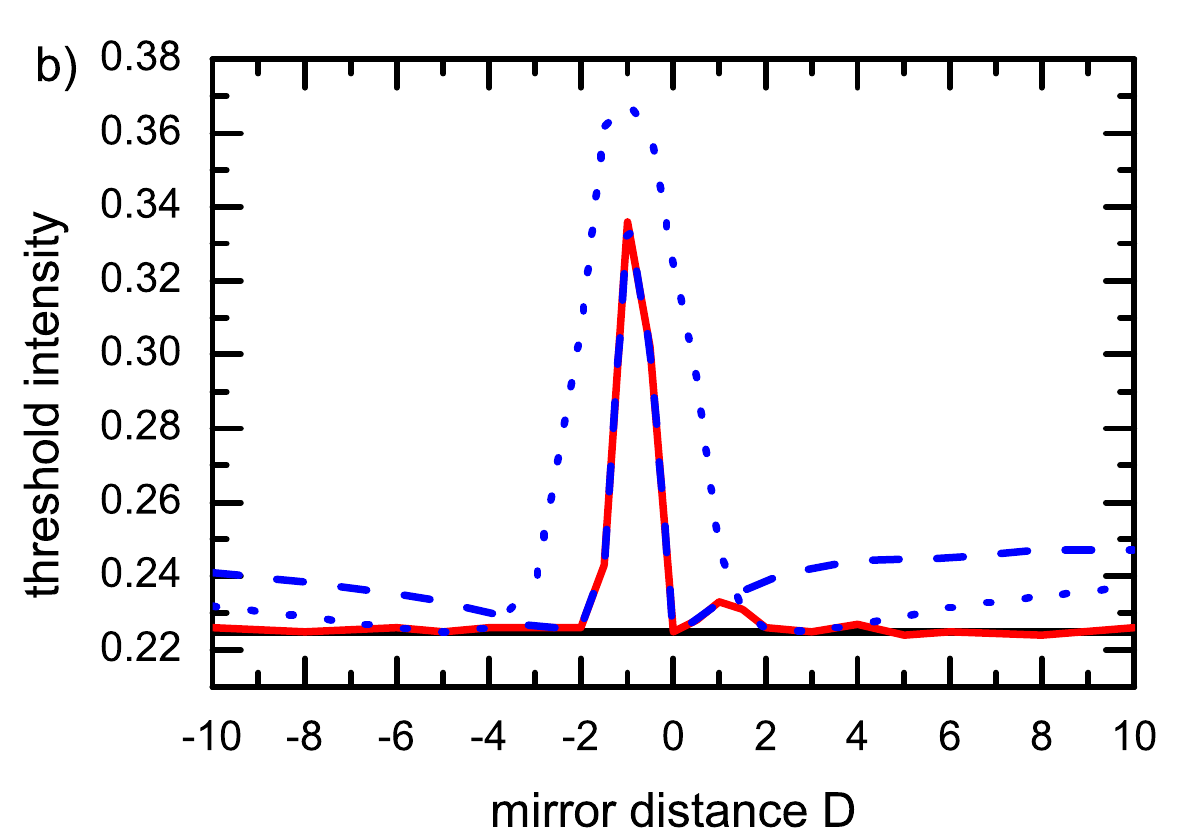}
\caption{(Color online) a) Pattern length scale (characterized by diffraction parameter $\theta$) and b) threshold intensity vs mirror distance $D$ for a self-focusing Kerr medium with $h=1$ described by $\hat{F}_{Kerr}$. Red solid curve: minimum threshold, blue dashed curve: lowest wavenumber (first Talbot) balloon, blue dotted curve:  second lowest wavenumber (second Talbot) balloon, black solid curve: minimum threshold condition for CP instability, Fig.~\ref{fig:KerrG2D0}.}

 \label{fig:D_thres}
 \end{figure}

Another way to illustrate this point is visualized in Fig.~\ref{fig:D_thres}.  The red solid curve in Fig.~\ref{fig:D_thres}a denotes the length scale of the minimum threshold mode vs mirror distance. For $D=-3 \ldots 1$ it mirrors the first Talbot balloon, until it jumps to the second and follows it for $D=-6 \ldots -4$ and $D=1.5 \ldots 4$. Afterwards it jumps again and wiggles around a horizontal, which is very close to the value for the CP instability at twice the medium length or the SFM instability at $D=0$ (Fig.~\ref{fig:KerrG2D0}). The changes of lengthscale imply that the minimum of the envelope curve is at finite $\theta$ and the system is trying to stay close to this value as far as compatible with the specific boundary conditions, i.e. diffractive phase shift $\theta$ at the feedback distance $D$. These considerations are maybe even more apparent for the thresholds (Fig.~\ref{fig:D_thres}b) where the SFM and CP threshold curves are nearly indistinguishable at large $|D|$.

\section{Saturable absorption and approach to atomic resonance}

As mentioned, the quasi-Kerr treatment of the atomic susceptibility is valid only for large atomic detunings.  Nonlinear effects typically strengthen as detuning is decreased and atomic resonance is approached, but resonant absorption kills the feedback. It is therefore important to extend our models to address the absorptive response at finite detunings, which implies using  z-dependent forward and backward intensities in the transverse perturbation problem. From the structure of (\ref{transpertK}), it is evident that the presence of the diffraction parameter $\theta$ mixes the real and imaginary parts of the perturbations ($f,g$) and thus adds significant mathematical complication. As a first approach to inclusion of absorption, therefore, it is worthwhile to analyze the case in which $\theta$ is set equal to zero. Physically, this corresponds to neglecting diffraction within the medium, often referred to as the "thin medium" approximation. As well as linking to work in which the medium is regarded as a thin slice, this approach also enables consideration of multi-slice models \cite{firth90b}, where the medium is approximated by a sequence of slices with free-space diffraction in-between. Split-step numerical algorithms typically adopt such an approach, so there is also computational interest in this approximation.

In our earlier discussion of the envelope functions in the quasi-Kerr approximation, we saw that the thin-slice limit $\theta=0$ is typically approached with finite slope, but nevertheless with a  threshold of the same order as those found for optimum mirror distances. We can therefore expect that the thin-medium approximation will offer a worthwhile qualitative picture of the effect of linear and nonlinear absorption on thresholds and tuning ranges as atomic resonance is approached, and indeed we will find behaviors in rather good agreement with the cold-atom patterns reported in \cite{Camara2015}.

Dropping diffraction  and assuming threshold conditions, the system (\ref{transpert}) becomes:
\begin{equation}
\label{thinpert}   \left \{
\begin{array}{l}
\frac{df}{dz}  = -\alpha_lL(1+i\Delta)(F_{11}f'+F_{12}g') ,\\
\frac{dg}{dz}  = \alpha_lL(1+i\Delta)(F_{21}f'+F_{22}g') \\
\end{array} \right.
\end{equation}

 In the presence of absorption, the elements of $\hat{F}$ are z-dependent, for example obeying the zero-order solutions derived above for  various models. Note that the imaginary parts of both $f$ and $g$ are slaved to the real parts. In particular, if $f(0)=f_0=0$, as for the input to a SFM system, then $f(1)=f_1=(1+i\Delta) f_1^{\prime}$. The usual mirror feedback conditions for a transverse perturbation then imply $dq_1= R(cos\psi_D+\Delta sin\psi_D)dp_1$, where $dp_1$ and $dq_1$ are the intensity changes associated with $f_1$ and $g_1$.

Instead of integrating the system  (\ref{thinpert}) we adopt a different approach. Since we have neglected diffraction in the medium, the perturbed system obeys the same equation as the homogeneous solution, but with perturbed boundary conditions. Specifically, for a given output $p_1$, and corresponding feedback $q_1 = R p_1$, we can analytically and/or numerically calculate the corresponding input $p_0$. Using the same algorithm, we can formally calculate the change in $p_0$ due to a small change $dp_1$ in $p_1$ with no change in $q_1$, and conversely. We can thus find the ratio of $dp_1$ to $dq_1$ which leaves $p_0$ unchanged to first order - which is the input boundary condition. Only for specific values of $p_1$ will this ratio be equivalent to the feedback phase relation $dq_1= R(cos\psi_D+\Delta sin\psi_D)dp_1$ identified above. Finding such $p_1$ values, and the corresponding input values $p_0$, gives pattern thresholds for the assumed values of $R$, $\Delta$, $\theta$ and $D$. We then eliminate $D$ and $\theta$ by requiring that that the perturbation gain is maximised, which implies $ cos\psi_D+\Delta sin\psi_D = \sqrt(1+\Delta^2)$. With these choices we find that the maximal pattern-forming region is a closed domain in the remaining parameter space ($p_0, \Delta$)  for both the no-grating and all-grating two-level models.

\begin{figure}

\includegraphics[scale=0.9,width=\columnwidth]{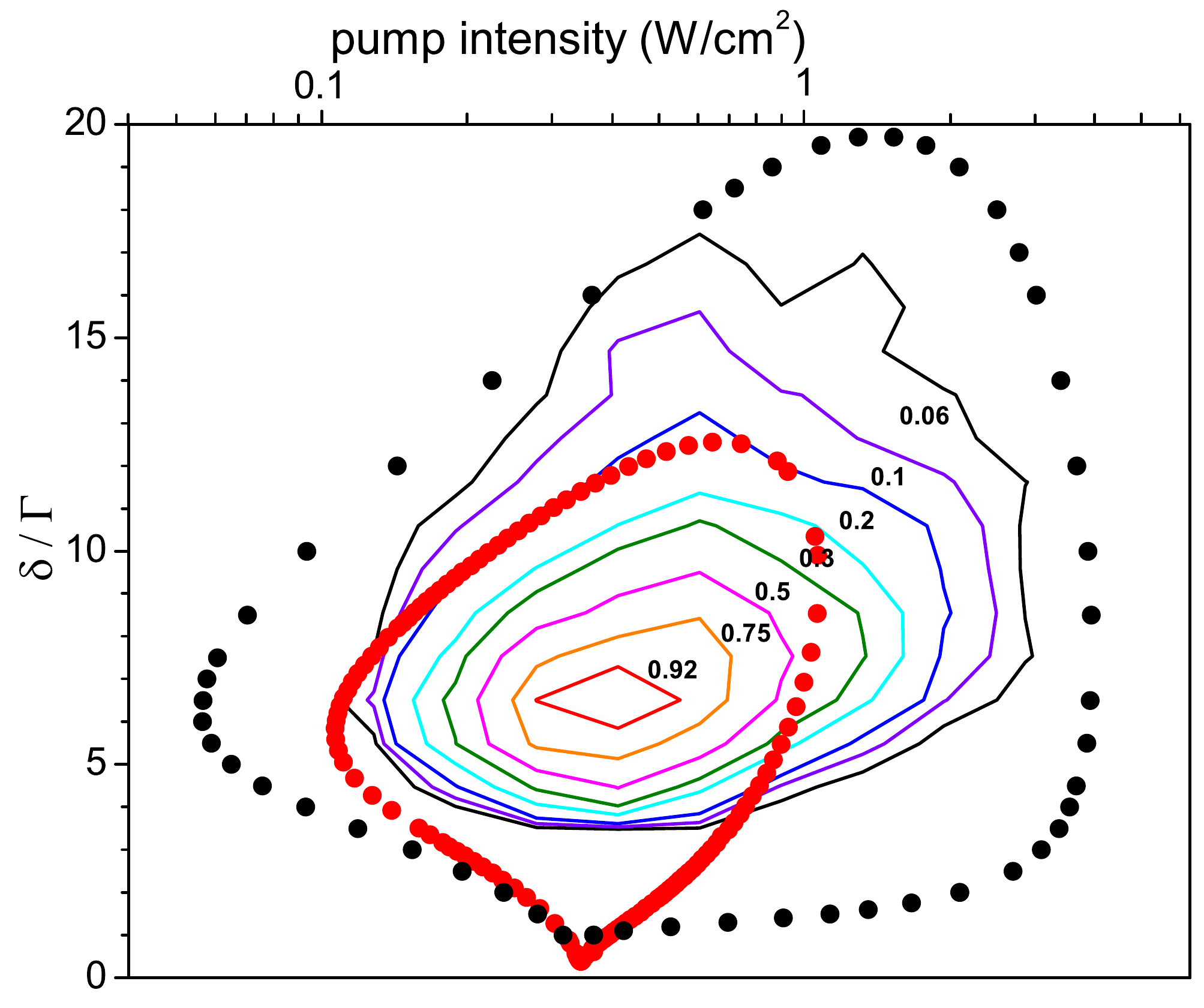}
\caption{ (Color online) (a) Two-level instability domain ($\delta > 0$) reported in \cite{Camara2015}.
Diffracted power  $P_d$  is measured as a function of $\delta > 0$ (note that $\Delta=2\delta/\Gamma$) and input intensity $I$, and  the data plotted
as isolines. Note the logarithmic horizontal scale. The dotted loops indicate maximal instability domains calculated in the thin-medium approximation as described in the text: (black) domain calculated from  (\ref{xi_0}), i.e. with all reflection gratings included ($h=1$);  (red) domain calculated from  (\ref{p_1nograt}), i.e. with no reflection gratings ($h=0$). Both dotted traces are rescaled to absolute values of intensity and detuning. }
 \label{CamaraTuning}
 \end{figure}

Figure~\ref{CamaraTuning} compares the threshold domains for these two thin-slice, all-tuning, absorptive models with  experimental data  \cite{Camara2015} on the detuning behavior of the diffracted power observed under pattern formation conditions in a cold Rb cloud with single feedback mirror.The agreement for the all-grating model is rather satisfactory, bearing in mind that the theory only calculates threshold conditions, while the experiment detects diffracted power only if the perturbation gain is large enough to build a strong pattern from noise within the microsecond or so duration of the pump pulse. Moreover, we note that the no-grating threshold domain is smaller than that in which transverse structure is observed. This provides firm evidence that reflection gratings are present in the cold-atom cloud, in agreement with expectations based on the inability of transport mechanisms to wash out susceptibility gratings at such low temperatures when such short input pulses are used.

%
%


\section{Conclusion}
\label{conclusions}

In this paper we have undertaken a largely analytic investigation of thresholds and lengthscales for pattern formation in a saturable two-level medium, optically-excited close to resonance from one side, and with a feedback mirror to reflect and phase-shift the light fields after they have traversed the medium. In that scenario, we have established a number of results, in encouraging agreement with recent experimental results in several cases. 

Perhaps our main result is that thresholds for the feedback mirror (SFM) configuration are in exact correspondence with one set of threshold curves for symmetrically-excited counterpropagation (CP) in a medium of twice the length, when the mirror plane in the SFM system is at the output of the medium. One important consequence of this is that SFM thresholds are significantly lower than CP thresholds in the same sample (e.g. cloud of cold atoms). Since large cold-atom clouds are difficult to produce, this can make the difference between observing well-developed patterns and failing to reach threshold at all.

While this scaling result is  derived for a saturable nonlinearity with absorption neglected, it is a consequence of parity symmetry in the CP system, and should hold in relation to any CP system with parity symmetry. Assuming that there is a stable zero-order solution of the CP system equations which exhibits parity symmetry, any perturbation eigenmode of the system must be either symmetric or anti-symmetric at the central symmetry plane. For an even mode, and also for the zero-order solution, one can replace the CP system with a perfect mirror at the symmetry plane without essential change to the equations or the solutions, and the CP/SFM scaling follows.  Hence in a wide class of nonlinear optical systems, the SFM system offers an approximately fourfold advantage in power over a CP configuration using the same medium (twofold reduction in pumping power, and approximately twofold reduction in threshold power).

There is a further advantage of the SFM system, in that the mirror location $D$ can be varied continuously over a wide range around and beyond the medium length. As well as allowing the observed pattern scale to be quasi-continuously varied, which is  at the very least useful for diagnostics, it is also found that the minimum threshold usually occurs for $D \ne 0$, essentially because a non-zero feedback phase allows optimum matching between forward and backward perturbation growth rates. We have considered, and compared to experiment, the "Talbot fan" characteristics which characterize the evolution of pattern scales as $D$ is varied, and explained observed sudden changes of scale in terms of mode competition in the neighborhood of the minimum possible (in $D$) threshold.

The additional degree of freedom offered by finite $D$ also implies an additional complexity in the analysis. We have shown, however, that thresholds are constrained by envelope curves to which the threshold curves are tangent, and along which they evolve as $D$ is varied. Hence important properties of the SFM system such as the minimum possible threshold, and the domains within which pattern formation is possible (or impossible) can be found, often analytically. Again, the envelope propery is likely to be general, even though we have derived it only in the quasi-Kerr limit, because it follows from the structure of the feedback boundary condition.

Importantly, the envelope functions enable a quantitative investigation of the limit $D/L \to\infty$, which correspond to diffraction in the medium being negligible compared to that in the feedback loop, i.e the thin-slice limit. We find that threshold values tend to precisely the thin-medium values, but with finite slope. As a consequence we have demonstrated that the degeneracy of the unstable modes predicted in thin-medium theory does not survive inclusion of finite medium length, even at lowest order.

Diffusive damping removing the degeneracy was introduced in the first treatments \cite{firth90a,dalessandro92} to  model carrier diffusion in semiconductors or elasto-viscous coupling in liquid crystals, which will make these media deviate from purely local Kerr media. In hot atom experiments \cite{grynberg94,Ackemann1994,ackemann95b} the thermal motion of the atoms, which can be modelled under appropriate conditions \cite{Ackemann1994,ackemann95b} as diffusive motion, will in tendency provide a stronger wash-out for transverse gratings at larger wavenumber and thus remove the degeneracy. In cold atoms this effect is not very strong and the finite medium thickness appears to be the main mechanism responsible for the emergence of a defined length scale \cite{labeyrie14, Camara2015}.

In the specific context of the two-level nonlinearity we have analyzed different models to take account of wavelength scale (reflection) gratings in the steady-state susceptibility applicable to counterpropagation problems. We have found that models in which only the lowest-order (2k) gratings are considered predict a zero-order bistability as resonance is approached. This bistability disappears when all orders (m$\times$2k) of gratings are included, and is therefore probably spurious. We have been able to develop models which include all grating orders, in particular in the quasi-Kerr and thin-medium limits, and have demonstrated reasonable agreement with experiment using these all-grating models.

In summary, we have developed a firm and systematic foundation for the analysis of the effects of in-medium diffraction, and of reflection gratings, in SFM  pattern formation. Though we have focused here on the saturable  two-level electronic nonlinearity, our approach and techniques have applicability across a wide class of nonlinearities. While our present analysis deals only with thresholds and steady-state instabilities, these are an important, and even essential, preliminary to more extensive numerical simulations, necessarily involving many additional parameters and many spatial and temporal scales. We already showed \cite{labeyrie14} that a simple thick-medium Kerr model gives useful insight into optomechanical SFM patterns, and in this work we have shown that a similar analysis helps understand important features of polarization-mediated SFM patterns in cold atoms.  Patterns in cold-atom clouds with laser irradiation and mirror feedback are proving to a be a very rich field, with diverse implications, and a secure basis for the interpretation of experimental results and the development of appropriate theoretical models is therefore very important.

\acknowledgments{The Strathclyde group is grateful for support by the Leverhulme Trust and an university studentship for IK by the University of Strathclyde. The Sophia Antipolis group is supported by CNRS, UNS, and R\'{e}gion PACA. The collaboration between the two groups was supported by Strathclyde Global Exchange Fund and CNRS. WJF also acknowledges sharing of unpublished work by M. Saffman. We are grateful to A. Arnold and P. Griffin for experimental support, to G.R.W. Robb, G.-L. Oppo and R. Kaiser for fruitful discussions.}

\bibliography{bibl_FBM2level}

\vspace{\parindent}

\clearpage
\newpage
\section{Appendix}
In this Appendix we present a matrix approach to the analytic solution of (\ref{transpertK}) leading to the threshold formulae (\ref{2LSslab},\ref{2LSfbm}) for the CP and SFM problems respectively in the quasi-Kerr case. Our methods and results are broadly similar to those of  \cite{firth90b,Geddes1994}, but because of slight notational differences, and our extension to more general nonlinearities and the SFM problem, it is perhaps worthwhile to present the details of the analysis.\\

We analyze the system in terms of a real 4-component vector $U = [f',g',f'',-g'']^{tr}$, which obeys
\begin{equation}
\label{dUdz}
\frac{dU} {dz} = M U.
\end{equation}
where $M$ is a real $4 \times 4$ matrix with constant coefficients:
\begin{equation}
M=
\left(
\begin{array}{cccc}
 0  & 0 & \theta & 0   \\
  0 & 0  & 0 & \theta  \\
  -\theta-\kappa F_{11} &  -\kappa F_{12} & 0 & 0  \\
   -\kappa F_{21} & -\theta-\kappa F_{22}&0 &  0
\end{array}
\right) \nonumber
\end{equation}
\\ Here $\kappa =  \alpha_{l}L\Delta $ is an effective Kerr coefficient. \\

The formal solution to
(\ref{dUdz}) is
\\

$U(1) = exp(M) U(0)$ or $U(0) = exp(- M) U(1)$.\\
\\

For both CP and SFM cases, $f(0) =0$ is assumed, giving two conditions on the solution. The boundary
conditions at $z=1$ provide the necessary two additional equations. For the CP case, with input fields
at both ends, this condition is simply $g(1)=0$. For the feedback mirror case, however, the condition is that $f=g$ on the mirror, and hence $g(1) = exp(-2i\psi_D) f(1)$, where $\psi_D =D \theta/L$ governs the phase shift of the perturbation field in propagating an effective distance $DL$ to the mirror. The relative mirror distance $D$ can be negative if the feedback optics involves a telescope.
For both types of boundary condition the solution to (\ref{dUdz}) leads to a pair of homogeneous linear
equations for ($g'(0),g''(0)$) which have a non-trivial solution only if the determinant of the coefficients vanishes.
This condition determines the pattern formation threshold as a function of $Q^2$ and system parameters. Hence, given $expM$, the  quasi-Kerr limit is fully solvable for all the two-level models we have discussed,  for both the CP and SFM cases.

The problem thus hinges on exponentiation of the matrix $M$. It has has a similar form to that analysed in the Appendix to \cite{firth90b}, and can be analytically exponentiated in a similar fashion. Squaring $M$, we obtain a block-diagonal matrix, its diagonal submatrices both being $-C$, where the $2 \times 2$ matrix $C$ is given by $C=\theta(\theta +\kappa \hat{F}$).  The eigenvalues of $C$ are given by the parameters $\psi_{i}^2 = \theta(\theta +\kappa \phi_{i})$ introduced in the main text, where the $\phi_{i}$ are the eigenvalues of $\hat{F}$. It follows that any unitary transformation that diagonalizes $\hat{F}$ also diagonalizes $C$, which provides one route to calculation of $expM$. As mentioned above, for equal intensities the eigenvectors of  $\hat{F}$ are proportional to $(1, \pm 1)$, which enables an intensity-independent transformation on $(f,g)$ leading to explicit expressions for $expM$ (and $exp(Mz)$) in terms of the $\psi_i$, equivalent to those obtained in \cite{Geddes1994}. Because the SFM boundary conditions are more involved than the CP ones, and also to enable consideration of mirror reflectivity $R \ne 1$, we choose to use the ($f,g$) basis described by $U$.
\\

Because $M^2$ is block diagonal, we write
\begin{equation}
exp(M)= 1+ \frac{M^2}{2!} + \frac{M^4}{4!} + .... + M (1+ \frac{M^2}{3!} + \frac{M^4}{5!} + ....)
 \nonumber
\end{equation}
The power series in $M^2$ can be expressed as block-diagonal cosine and sinc functions of $\sqrt{C}$, a 2x2 matrix obeying $(\sqrt{C})^2 = C$. As in \cite{firth90b}, we can then write an explicit expression for $expM$ as a $2 \times 2$ block matrix:

\begin{equation}
\label{expM}
exp(M)=
\left(
\begin{array}{cccc}
 cos\sqrt{C}  & \theta  sinc\sqrt{C}  \\
 -( C / \theta) sinc\sqrt{C} & cos\sqrt{C}
\end{array}
\right)
\end{equation}.\\

Because the cosine and sinc are even functions, this expression for $exp(M)$ is unique in terms of $C$, even though $\sqrt{C}$ is not uniquely defined.\\

Suppose that the $2 \times 2$ matrix $E$ diagonalizes $\hat{F}$, i.e. $E\hat{F}E^{-1} = diag(\phi_1,\phi_2$). Then E also diagonalizes $C$, as $ diag(\psi_1^2,\psi_2^2)$, and hence any matrix function of $C$, such as those occurring in $exp(M)$. Defining $E_2$ as a diagonal $2 \times 2$ block matrix with $E$ as its diagonal blocks,  some manipulation readily leads to
\begin{equation}
\label{explicitM}
E_2 U(1)=
\left(
\begin{array}{cccc}
 c_1  & 0 & \theta s_1/\psi_1 & 0   \\
  0 & c_2  & 0 & \theta s_2/\psi_2  \\
  -\psi_1 s_1/\theta &  0 & c_1 & 0  \\
   0 & -\psi_2 s_2/\theta &0 &  c_2
\end{array}
\right) E_2 U(0)
\end{equation}
where $c_i = cos \psi_i$ and $s_i = sin \psi_i$. A similar equation holds for $U(z)$ at any position $0< z<1$ within the medium, with the arguments of the sines and cosines replaced by $\psi_i z$, so the evolution of the perturbations within the medium can also be calculated.\\

This analytic solution can be applied to any quasi-Kerr "slab" system, for any boundary conditions, whether CP or SFM, including the unequal intensity case $p \ne q$ (e.g. $R \ne 1$ for SFM). It can also be used to calculate probe gain, for example, i.e. for non-zero input perturbations.\\

Here we will only consider equal intensities, for which, as mentioned in the main text, the eigenvectors of $\hat{F}$ are simply given by $(1,\pm1)$, leading to a simple explicit expression for $E$:
\begin{eqnarray}
\label{Esym}
& & E= \frac{1}{\sqrt{2}}  \left(
\begin{array}{cccc}
 1 & -1 \\
1 & 1
\end{array}
\right). \nonumber
\end{eqnarray}

For the case of  counterpropagating inputs with the usual boundary conditions $f(0) = g(1) =0$, (\ref{explicitM})  leads, after some algebra, to the usual CP threshold formula (\ref{2LSslab}). Since $E$ is a simple constant matrix independent of any system parameters, one can conveniently consider $EU$ as a change of variables in  (\ref{explicitM}), which is effectively the approach of Geddes et al \cite{Geddes1994}.

For the $R=1$ feedback mirror, the right side of  (\ref{explicitM})  is the same as for the CP problem ($f(0)=0$), but the left side needs to express the feedback-phase relationship between $f(1)$ and $g(1)$. Using the appropriate boundary conditions leads to the threshold expression (\ref{2LSfbm})  in the main text. \\

For unequal intensities, the CP threshold expression was presented in \cite{firth90b}. It leads to an interesting phenomenon whereby the crossings of the two threshold curves $H_1=0$ and $H_2=0$ become anti-crossings, with oscillatory solutions along a line of Hopf bifurcation joining the static threshold curves. Because the SFM problem is not parity-symmetric, no such scenario exists in the $R \ne 1$ feedback mirror situation, and only quantitative effects on the threshold are expected.
\end{document}